\RequirePackage{amsmath}
\documentclass{tlp}
\usepackage{amssymb}
\usepackage{xcolor}
\usepackage{proof}

\usepackage{etoolbox}
\usepackage{mathtools}

\newcounter{listcount} \newcounter{totcount}

\newcommand{\printarray}[2][2.5em]{
  \unskip \setcounter{totcount}{0}
  \renewcommand*{\do}[1]{\stepcounter{totcount}}
  \docsvlist{#2}
  \setcounter{listcount}{0}
  \renewcommand*{\do}[1]{%
    \stepcounter{listcount}
    \framebox[#1][c]{\rule{0pt}{1.5ex}\smash{\ensuremath{##1}}}%
    \ifnum\value{listcount}<\value{totcount}\thickspace\fi
  }
  \docsvlist{#2}
}

\newcommand\bcmdtab{\noindent\bgroup\tabcolsep=0pt%
  \begin{tabular}{@{}p{10pc}@{}p{20pc}@{}}}
\newcommand\ecmdtab{\end{tabular}\egroup}

\title[Property-based testing for Spark Streaming]
      {Property-based testing for Spark Streaming}

 \author[A. Riesco and J. Rodr\'iguez-Hortal\'a]
        {A. RIESCO\\
         Universidad Complutense de Madrid, Spain
         \and J. RODR\'IGUEZ-HORTAL\'A\\
         \email{ariesco@fdi.ucm.es juan.rodriguez.hortala@gmail.com}
         }

\jdate{September 2001, update April 2007}
\pubyear{2001}
\pagerange{\pageref{firstpage}--\pageref{lastpage}}
\doi{S0956796801004857}

\newtheorem{lemma}{Lemma}[section]
\newtheorem{definition}{Definition}[section]
\newtheorem{example}{Example} 
\newtheorem{theorem}{Theorem} 

\usepackage{url}
\usepackage{hyperref}

\newcommand{\codesize}{\small}

\newcommand{\lola}{\textsc{Lola}}

\newcommand{\mi}[1]{\mathit{#1}}
\newcommand{\nt}{\mathit{nt}}
\newcommand{\true}{\top}
\newcommand{\false}{\bot}

\newcommand{\nc}{\newcommand}
\nc{\parts}[1]{\mathcal{P}({#1})} 
\nc{\sstl}{\mathit{LTL_{ss}}} 
\nc{\tuntil}[1]{~U_{#1}~} 
\nc{\next}{X}
\nc{\teventually}[1]{\lozenge_{#1}} 
\nc{\always}[1]{\square_{#1}}
\nc{\trelease}[1]{~R_{#1}~} 
\nc{\now}[2]{\lambda^{#2}_{#1}}

\nc{\term}{{\it Term}}
\nc{\pred}{\mathcal{P}}
\nc{\fs}{\mathcal{F}}
\nc{\var}{\mathcal{V}}
\nc{\letter}{\Sigma} 
\nc{\timed}{\mathbb{N}}
\nc{\form}{{\sstl}} 
\nc{\sed}{{\it A}} 
\nc{\sei}{{\it I}} 
\nc{\ase}{\mathcal{A}} 
\nc{\sse}{\mathcal{A}^{\mathcal{S}}} 
\nc{\jud}[4]{{#1} \vDash^{#4} {#2} : {#3}} 
\nc{\juda}[3]{{#1} \vDash^{\ase} {#2} : {#3}} 

\newcommand{\den}[1]{[\![#1]\!]}

\begin{document}

\label{firstpage}

\maketitle

\begin{abstract}
Stream processing has reached the mainstream in the last years, 
as a new generation of open source distributed stream processing systems, 
designed for scaling horizontally on commodity hardware, has brought the capability 
for processing high volume and high velocity data streams to companies of all sizes. 
%
In this work we propose a combination of temporal logic and property-based testing (PBT)
for dealing with the challenges of testing programs that employ this programming model. 
We formalize our approach in a discrete time 
temporal logic for finite words, with some additions to improve the expressiveness of 
properties, which includes timeouts for temporal operators and a binding operator for letters. 
In particular we focus on testing Spark Streaming programs written with the Spark API for the functional
language Scala, using the PBT library ScalaCheck. 
For that we add temporal logic operators to a set of new ScalaCheck generators and properties, as part 
of our testing library sscheck. 
Under consideration in Theory and Practice of Logic Programming (TPLP).

\end{abstract}


\section{Introduction}\label{sec:intro}
%

With the rise of Big Data technologies \cite{marz2015big}, distributed 
stream processing systems (SPS)
\cite{akidau2013millwheel,marz2015big} have gained popularity in the last years. 
This later generation of SPS systems, characterized by a distributed architecture designed for horizontal scaling,  was pioneered by Internet-related companies, that had to find innovative solutions to scale their systems to cope with the fast growth of the Internet.
These companies are able to continuously process high volume streams of data by using systems like MillWheel~\cite{akidau2013millwheel}, Apache Storm~\cite{marz2015big}, Heron \cite{twitter2015Heron}, S4 \cite{neumeyer2010s4}, and Samza~\cite{gorawski2014survey}.
However, the first precedents of stream processing systems come back as far as the
early synchronous 
data-flow programming languages like Lutin~\cite{lutin} or 
Lustre~\cite{halbwachs1992synchronous}. 
%
A plethora of new distributed SPS have arisen in the last years, with 
proposals like Apache Flink~\cite{carbone2015apache}, Akka Streams~\cite{kuhn2014reactive}, and Spark Streaming~\cite{zaharia2013discretized}.
%
%
Among them Spark Streaming stands out as a particularly popular option in the industry. A basic indicator for that is the evolution of the search terms for different SPS on Google Trends, showing Spark Streaming as the most popular SPS from January 2016 onwards \cite{googleTrendsSpark2018}. 
%
%
%
Spark \cite{zaharia2012resilient} is a distributed processing engine designed for processing large collections of data \cite{white2012hadoop}. The core abstraction of Spark is the notion of Resilient Distributed Dataset (RDD), which provides a fault tolerant implementation of distributed collections. 
Spark Streaming is a SPS built on top of Spark, and it is based on the notion of  DStreams (Discretized Streams), which are series of RDDs corresponding to splitting an input data stream into fixed time windows called micro-batches, that are generated at a fixed rate according to a configured \emph{batch interval}. Spark Streaming is synchronous in the sense that the batches for all DStreams in the program are generated at the same time, as the batch interval is met. See \ref{sect:spark-intro} for a quick introduction to Spark and Spark Streaming.

In this work we present a framework to test temporal properties on stream-processing systems. Among
others, we consider that \emph{safety} (something bad never happens) and \emph{liveness} (something 
good eventually happens) properties are especially relevant in this kind of systems and might be 
useful for programmers. 
Specifically, we explore the problem of developing a testing library for Spark Streaming. 
We focus on Spark Streaming because its popularity implies a bigger set of potential users for our system, and in particular on its Scala API.
We are interested in logic-based approaches, that nevertheless can be applied by software developers without the assistance of formal verification experts, as part of a test-driven development (TDD) cycle \cite{beck2003test}. 
Testing an SPS-based program is intrinsically hard, because it requires handling time and events. 
%
%
Among the different proposals in the literature that tackle the problem of testing and modeling systems that deal with time, we follow  Pnueli's approach~\cite{pnueli1986applications}, that uses temporal logic for testing reactive systems. 
We define the logic $\sstl$, a variant of first-order linear temporal logic (LTL) \cite{modalLogic} that is suitable for expressing Spark Streaming computations, which we expose to programmers as the sscheck library \cite{sscheckSW}. This library extends ScalaCheck~\cite{nilsson2014scalacheck}, a popular property-based testing (PBT)~\cite{claessen2011quickcheck} library for Scala. 
In ScalaCheck a test is expressed as a property, which is a formula in a restricted version of first-order logic that relates program inputs and outputs. Each quantifier in the formula is bound to a generator function, that randomly produces values for some data type. The testing framework checks the property by evaluating it against a specified number of inputs that are produced by the generators.
%
That provides a sound procedure for checking the validity of the formulas implied by the properties, that is not complete but that it is fast and lightweight enough to integrate in a TDD cycle---see \ref{sect:pbt-intro} for an overview of PBT and ScalaCheck.  
sscheck extends ScalaCheck with temporal logic operators that can be used both in the generator functions for the input DStreams, and in the quantified formula that relates input and output DStreams, which simplifies expressing complex conditions on the sequence of batches for each DStream. 

As DStreams are discrete, $\sstl$ uses \emph{discrete time}. Our logic also considers {finite words}, like those used in the field of runtime verification~\cite{leucker2009brief}, instead of infinite $\omega$-words as usual in model checking. That allows us to easily integrate with the simple property checking mechanism of ScalaCheck. 
%
%
Although Spark DStreams are supposed to run indefinitely, so we might had modeled each DStream by an infinite word, in our setting we only model a finite prefix of the DStream. This 
allows us to implement a simple, fast, and sound procedure for evaluating test cases, based on evaluating the property on the generated finite prefix. 
On the other hand the procedure is not complete because only a prefix of the DStream is evaluated, but anyway PBT was not complete in the first place. Hence a \emph{test case} will be a tuple of \emph{finite prefixes} of DStreams, which corresponds to a \emph{finite word} in this logic, and the aforementioned external quantifier ranges over the domain of finite words. 
%
We provide a precise formulation for our logic $\sstl$ in Section \ref{sec:bdltl},
while details on how to implement properties are presented in Section~\ref{sect:implementation-details},
but for now let's consider a concrete example in order to get a quick grasp of our proposal. 
\begin{example}\label{example-banning}
We would like to test a Spark Streaming program that receives a stream of events describing user activity in some system. The program outputs a stream with the identifiers of banned users, which are users that the system has detected as abusing the system based on their activity. 
To keep the example simple, 
we assume that the input records are pairs containing a \texttt{Long} user id, and a \texttt{Boolean} value indicating whether the user has been honest at that instant. The output stream should 
include the ids of all those users that have been malicious now or in a previous instant. 
So, the test subject that implements it has type \texttt{testSubject: DStream[(Long, Boolean)] => DStream[Long])}. 
%

%

To define a property that captures the expected behavior, we start by using sscheck to define a \emph{generator} for (finite prefixes of) the input stream. As we want this input to change with time, we use a temporal logic formula to specify the generator. 
We start by defining the atomic non-temporal propositions, which are generators of micro batches with type \texttt{Gen[Batch[(Long, Boolean)]]}, where \texttt{Batch} is a class extending \texttt{Seq} that represents a micro batch. We can generate good batches, where all the users are honest, and bad batches, where a user has been malicious. We generate batches of \texttt{20} elements, and use \texttt{15L} as the id for the malicious user: 
%

{\codesize
\begin{verbatim}
val batchSize = 20 
val (badId, goodIds) = (15L, Gen.choose(1L, 50L))   
val goodBatch = BatchGen.ofN(batchSize, goodIds.map((_, true)))
val badBatch = goodBatch + BatchGen.ofN(1, (badId, false))
\end{verbatim}
}

\noindent
where \texttt{BatchGen.ofN} is a function that generates a batch with the specified number of elements, that are generated by the generator function in its second argument. 
%
In our logic $\sstl$, that corresponds to some predicate symbols under an interpretation structure $\ase = (\sed, \sei)$ where the domain $\sed$ is the set of all Scala expressions. As we are verifying a Spark Streaming program, given a $\mathbb{U} = \mathbb{RDD} \times \mathbb{RDD}$ for $\mathbb{RDD}$ the set of all Spark RDDs, we use timed  letters in $\mathbb{U}^t = \mathbb{U} \times \mathbb{N}$
that correspond to an RDD for the input and another for the output DStream, together with the time at which the letter is produced. The interpretation function $\sei$ would be defined as follows, assuming the usual meaning for the cardinal ($\#$) and inclusion ($\in$) operators for $\mathbb{RDD}$, and identifying Scala number literals with the corresponding numbers by abuse of notation.

{\codesize
$$
\begin{array}{l}
\mathit{goodId} = \{ x \in \mathbb{Z} ~|~ 1 \leq x \leq 50 \}  
~~~\mathit{goodElem} = \{ e \in \mathbb{Z} \times \{\true, \false\} ~|~ e = (x, \true) \wedge x \in \mathit{goodId} \} \\
\sei(\mathit{goodBatch}) = \{ ((rdd_i, rdd_o), t) \in \mathbb{U}^t ~|~ \#(rdd_i) = 20 \wedge \forall e \in rdd_i.~ e \in \mathit{goodElem} \} \\
\sei(\mathit{badBatch}) = \{ ((rdd_i, rdd_o), t) \in \mathbb{U}^t ~|~  \#(rdd_i) = 20 \\
~~~~~~~~~~~~~~~~~~~~~~~~~~~~~~~~~~~~~~ \wedge \forall e \in rdd_i.~ e \in \mathit{goodElem} \vee e = (15, \false) \} \\
\end{array}
$$
}

So far generators are oblivious to the passage of time. But in order to exercise the test subject thoroughly, we want to ensure that a bad batch is indeed generated, and that several arbitrary batches are generated after it, so we can check that once a user is detected as malicious, it is also considered malicious in subsequent instants. Moreover, we want all this to happen within the confines of the generated finite DStream prefix. This is where \emph{timeouts} come into play. In our temporal logic we associate a timeout to each temporal operator, that constrains the time it takes for the operator to resolve. For example in a use of \emph{until} with a timeout of $t$, the second formula must hold before $t$ instants have passed, while the first one must hold until that moment. Translated to generators this means that in each generated DStream prefix a batch for the second generator is generated before $t$ batches have passed, i.e.\ between the first and the $t$-th batch. This way we facilitate that the interesting events had enough time to happen during the limited fraction of time considered during the evaluation of the property:
%

{\codesize
\begin{verbatim}
val (headTimeout, tailTimeout, nestedTimeout) = (10, 10, 5) 
val gen = BatchGen.until(goodBatch, badBatch, headTimeout) ++ 
          BatchGen.always(Gen.oneOf(goodBatch, badBatch), tailTimeout)
\end{verbatim}
}

The resulting generator \texttt{gen} has type \texttt{Gen[PDStream[(Long, Boolean)]]}, where \texttt{PDStream} is a class that represents sequences of micro batches corresponding to a DStream prefix. Here \texttt{headTimeout} limits the number of batches before the bad batch occurs, while \texttt{tailTimeout} limits the number of arbitrary batches generated after that. 
That generator corresponds to the  $\sstl$ formula below, that is defined employing versions with timeout of classical temporal logic operators like $\always{t}$ (always for $t$ batches), or $\tuntil{t}$ (until
for $t$ batches), as well as a new ``consume'' operator $\now{x}{t}$ that is basically a variant of the classical ``next'' operator that binds a variable to the current letter and time in the input word. 

{\codesize
$$
\begin{array}{l}
(\now{rdds}{o}.~\mathit{goodBatch}(rdds) \tuntil{10} \now{rdds}{o}.~\mathit{badBatch}(rdds)) \tuntil{100} \\
(\always{10}.~\now{rdds}{o}.~\mathit{goodBatch}(rdds) \vee \mathit{badBatch}(rdds))
\end{array}
$$
}

The output DStream is the result of applying the test subject to the input stream. 
We define the \emph{assertion} that completes the property as a 
\emph{temporal logic formula}:

{\codesize
\begin{verbatim}    
type U = (RDD[(Long, Boolean)], RDD[Long])
val (inBatch, outBatch) = ((_: U)._1, (_: U)._2)
val formula = {
  val allGoodInputs = at(inBatch)(_ should foreachRecord(_._2 == true))
  val badInput = at(inBatch)(_ should existsRecord(_ == (badId, false)))
  val noIdBanned = at(outBatch)(_.isEmpty)
  val badIdBanned = at(outBatch)(_ should existsRecord(_ == badId))
    
  ((allGoodInputs and noIdBanned) until badIdBanned on headTimeout) and
  (always { badInput ==> (always(badIdBanned) during nestedTimeout) }
           during tailTimeout)  }  
\end{verbatim}
}

Atomic non-temporal propositions correspond to assertions on the micro batches for the input and output DStreams. 
We use a syntax where the function \texttt{at} below is used with a projection function like \texttt{inBatch} or \texttt{outBatch} to apply an assertion on part of the current letter, e.g.\ the batch for the current input. 
The assertions \texttt{foreachRecord} and \texttt{existsRecord} are custom Specs2 assertions that allow
users to check whether a predicate holds or not for all or for any of the records in an RDD, respectively. 
This way we are able to define non-temporal atomic propositions like \texttt{allGoodInputs}, that states that all the records in the input DStream correspond to honest users.
%
%
But we know that \texttt{allGoodInputs} will not be happening forever, because \texttt{gen} eventually creates a bad batch, so we combine the atomic propositions using temporal operators to state things like ``we have good inputs and no id banned until we ban the bad id'' and ``each time we get a bad input we ban the bad id for some time.'' Here we use the same timeouts we used for the generators, to enforce the formula within the time interval where the interesting events are generated. Also, we use an additional \texttt{nestedTimeout} for the nested \texttt{always}. Timeouts for operators that apply an universal quantification on time, like \emph{always}, limit the number of instants that the quantified formula needs to be true for the whole formula to hold. In this case we only have to check \texttt{badIdBanned} for \texttt{nestedTimeout} batches for the nested \texttt{always} to be evaluated to true. 
That corresponds to the following $\sstl$ formula $\varphi$, assuming the interpretation of the predicate symbols $\mi{allGoodInputs}$, $\mi{badInput}$, $\mi{empty}$, and $\mi{badIdBanned}$ as specified below.

{\codesize
$$
\begin{array}{l@{\!\!\!\!\!\!\!\!\!\!\!\!\!\!\!\!\!\!\!\!\!\!\!\!\!\!\!}l}
\sei(\mathit{allGoodInputs}) = \{ (rdd_i, rdd_o) \in \mathbb{U} ~|~ \forall e \in rdd_i.~ e = (x, \true) \} \\
\sei(\mathit{badInput}) = \{ (rdd_i, rdd_o) \in \mathbb{U} ~|~ \exists e \in rdd_i.~ e = (15, \false) \} \\
\sei(\mathit{noIdBanned}) = \{ (rdd_i, rdd_o) \in \mathbb{U} ~|~ \not \exists e \in rdd_o\}  \\
\sei(\mathit{badIdBanned}) = \{ (rdd_i, rdd_o) \in \mathbb{U} ~|~ 15 \in rdd_o\} \\
(\now{rdds}{o}.~\mathit{allGoodInputs}(rdds) \wedge \now{rdds}{o}.~\mathit{noIdBanned}(rdds)) \tuntil{10}  \now{rdds}{o}.~\mathit{badIdBanned}(rdds)~~~~~~~~~~~~~~ \\
\wedge~ \always{10}.~\now{rdds}{o}.~\mathit{badInput}(rdds)  \rightarrow (\always{5}.~\now{rdds}{o}.~\mathit{badIdBanned}(rdds))
\end{array}
$$
}


Finally, we use our temporal universal quantifier \texttt{forAllDStream} to put together the temporal generator and formula, getting a property that checks the formula for all the finite DStreams prefixes produced by the generator:
 
{\codesize
\begin{verbatim}    
  forAllDStream(gen)(testSubject)(formula).set(minTestsOk = 20)
\end{verbatim}
}

The property fails as expected for 
a faulty stateless implementation that is not able to remember which users had been malicious in the past, and succeeds for 
a correct stateful implementation (see 
\cite{tplpTR} for details).
%
\end{example}

We carried out these ideas on the library sscheck \cite{sscheckSW}, previously presented in the tool paper \cite{sscheckIFM}, and in a leading engineering conference \cite{riesco2016apacheTemporal}. 
Moreover, sscheck has been discussed by others \cite{karauSparkTestStrata2015} and it
has also been referred in  books and technical blogs remarkable in the field \cite{KarauTestSpark2015b,karau2017high}, showing that it presents a good performance and
that it stands
as an alternative choice for state-of-the-art testing frameworks.

The present paper extends the published works above by:
\begin{itemize}
\item
Improving the logic by (i) redefining the semantics of formulas using a first order structure
for letters, that are evaluated under a given interpretation, (ii) introducing a new operator that allows us to bind the content
and the time in the current batch, (iii) redefining the previous results for the new logic,
and (iv) defining a new recursive definition that allows us to simplify formulas in a lazy way. 

\item
Formally proving the theoretical results arising from the new formulation.

\item 
Formalizing the generation of words from formulas. 


\item
Providing extensive examples of sscheck properties, including safety and liveness properties. 
\end{itemize}

The rest of the paper is organized as follows: 
Section~\ref{sec:logic} describes our logic for testing stream processing systems, while
Section~\ref{sec:tool} presents its implementation for Spark. Section~\ref{sec:rel}
discusses some related work.
Finally, Section~\ref{sec:conc} concludes and presents some subjects of future work.
The code of the tool, examples, and much more information is available 
in~\url{https://github.com/juanrh/sscheck}.
An extended version of this paper can be found in~\cite{tplpTR}.


\section{A Logic for Testing Stream Systems}\label{sec:logic}

We present in this section our linear temporal logic for defining properties on  
Spark Streaming programs. 
We first define the basics of the logic, 
then we show a stepwise formula evaluation procedure that is the basis for our
prototype, and finally we formalize the generation of test cases from formulas. 

\subsection{A Linear Temporal Logic with Timeouts for practical specification of stream processing systems \label{sec:bdltl}}

The basis of our proposal is the $\sstl$ logic, a linear temporal logic that 
combines and specializes both
$\mathit{LTL}_3$~\cite{bauer2006monitoring} and
First-order Modal Logic~\cite{fol-modal}, borrowing some ideas from TraceContract \cite{TraceContract}.
%
$\mathit{LTL}_3$ is an extension of $\mathit{LTL}$ \cite{reallyTemporalLogic} for runtime verification that takes into
account that only \emph{finite} executions can be checked, and hence a new value ?
(inconclusive) can be returned in case a property cannot be effectively evaluated to either
\emph{true} ($\true$) or \emph{false} ($\false$) in the given execution, because the word 
considered was too short. These values
form a lattice with $\false \leq\ ? \leq \true$.
$\sstl$ uses the same domain as $\mathit{LTL}_3$ for evaluating formulas, and the same truth tables for the basic non-temporal logical connectives ---see \cite{tplpTR} for details. 
$\sstl$ is also influenced by First-order Modal Logic, an extension to First-order of the
standard propositional modal logic approach~\cite{modalLogic}. 
Although the propositional
approach in~\cite{sscheckIFM} was enough for generating new values and dealing with some
interesting properties ---including safety properties--- we noticed that some other
properties involving variables bound in previous letters ---e.g. some liveness properties--- 
could not be easily specified in our logic. For this reason we have extended the original 
version of $\sstl$ with a binding operator inspired by a similar construction from TraceContract \cite{TraceContract}, 
which provides a form of universal quantification over letters, that makes it easy to define
liveness properties, as we will explain in Section~\ref{sect:twitter-example}. 
%
Note that timeouts for universal time quantifiers help relaxing the formula so its evaluation is conclusive more often, while timeouts for existential time quantifiers like \emph{until} make the formula more strict. We consider that it is important to facilitate expressing properties with a definite result, as quantifiers like \emph{exists}, that often lead properties to an inconclusive evaluation, have been abandoned in practice by the PBT user community \cite{nilsson2014scalacheck,venners2015Exists}.



\paragraph{Formulae Syntax}
We assume a denumerable set $\var$ of variables ($x, y, z, o, \ldots$), a denumerable set $\pred$ of predicate symbols ($p, q, r, \ldots$) with associated arity---with $\pred^n$ 
the set of predicate symbols with arity $n$, and $\mathbb{N} \subseteq \fs^0$---,
and a denumerable set $\fs$ of function symbols ($f, g, h, \ldots$) with associated arity---with 
$\fs^n$ 
the set of function symbols with arity $n$. Then, terms $e \in \term$ are built as:
$$
\begin{array}{ll}
\term \ni e ::= & x \mid f(e_1, \ldots, e_n) \text{ for } x \in \var, f \in \fs^n, e_1, \ldots, e_n \in \term \\
\end{array}
$$
Typically, propositional formulations of LTL \cite{reallyTemporalLogic} consider words that use the
power set of atomic propositions as its alphabet. However, we consider the alphabet 
$\letter = \term \times \timed$ of timed terms. 
Over this alphabet we define the set of finite words $\letter^*$, 
i.e. finite sequences of timed terms. We use $\epsilon$ for the empty word, and the notation $u = u_1 \ldots u_n$ to denote that $u \in \letter^*$
has length $\mathit{len}(u)$ equal to $n$, and $u_i$ is the letter at position $i$ in $u$. Each letter $u_i \equiv (e_i, t_i)$ corresponds to 
the term $e_i$ that can be observed at instant $i$ after $t_i$ units of time have been elapsed. For example, for 
a Spark Streaming program with one input DStream and one output DStream, the term $e_i$ would correspond to a pair
of RDDs, one representing the micro batch for the input DStream at time $t_i$, and another the micro batch 
for the output DStream at time $t_i$. 
~\\
It is important to distinguish
between the instant $i$, which refers to logic time and can be understood as
a ``counter of states,'' and $t_i$, which refers to real time. This real time
satisfies the usual condition of \emph{monotonicity}
($t_i \leq t_{i+1}, i \geq 0$), but does not satisfy \emph{progress}
($\forall t \in \mathbb{N}$, $\exists_{i \geq 0} t_i > t$), since we work with
finite words. It is also important to note that time is discrete but the time
between successive states may be arbitrary. 
Also note that by the condition $\mathbb{N} \subseteq \fs^0$, time literals are also terms, and therefore we can replaces variables by terms and still obtain a term, as we will do later on in this section when defining the semantics of formulas. 

%
%
%
%
The set of $\sstl$ formulas $\form$ is defined as follows: 
$$
\begin{array}{ll}
\form \ni \varphi ::= & \false \mid \true \mid p(e_1, \ldots, e_n) \mid e_1 = e_2 \mid 
               \neg \varphi \mid  \varphi_1 \vee \varphi_2 \mid
			   \varphi_1 \wedge \varphi_2 \mid \varphi_1 \rightarrow \varphi_2 \mid \\
            & \next \varphi \mid \teventually{t} \varphi \mid \always{t} \varphi 
              \mid \varphi \tuntil{t} \varphi \mid \varphi \trelease{t} \varphi
              \mid \now{x}{o}.\varphi
\end{array}
$$
We will use the notation $\next^n \varphi, n \in \mathbb{N}^+$, as a shortcut for $n$ applications
of the operator $\next$ to $\varphi$. Although we provide a precise formulation for the interpretation
of these formulas later in this section, the underlying intuitions are as follows: 
%
\begin{itemize}
\item The first eight formulas are based on classical first order non-temporal logical connectives, including
contradiction, tautology, atomic formulas based on predicate application and equality, and the negation 
and the usual binary connectives. 

\item
$\next \varphi$, read ``next $\varphi$'', indicates that the formula $\varphi$ should hold in the next state.

\item
$\teventually{t} \varphi$, read ``eventually $\varphi$ in $t$,'' indicates that
$\varphi$ holds in any of the next $t$ states (including the current one).

\item
$\always{t} \varphi$, read ``always $\varphi$ in $t$,'' indicates that
$\varphi$ holds in all of the next $t$ states (including the current one).

\item
$\varphi_1 \tuntil{t} \varphi_2$, read ``$\varphi_1$ holds until $\varphi_2$ in $t$,''
indicates that $\varphi_1$ holds until $\varphi_2$ holds in the next $t$ states,
including the current one, and $\varphi_2$ must hold eventually. Note that it
is enough for $\varphi_1$ to hold until the state previous to the one where
$\varphi_2$ holds.

\item
$\varphi_1 \trelease{t} \varphi_2$, read ``$\varphi_2$ is released by $\varphi_1$ in $t$,''
indicates that $\varphi_2$ holds until both $\varphi_1$ and $\varphi_2$ hold in the
next $t$ states, including the current one.
However, if $\varphi_1$ never holds and $\varphi_2$ always holds the formula holds
as well.

\item $\now{x}{o}.\varphi$, read 
``consume the current letter to produce $\varphi$'', indicates that 
given $(e,t)$ the letter for the current state, then the formula resulting from 
replacing in $\varphi$ the variables $x$ and $o$ by $e$ and $t$, respectively, should hold in the next state. We call this the \emph{consume} operator. 
\end{itemize}

We say that a formula is \emph{timeless} when it does not contain any of the temporal 
logical connectives. 
An $\sstl$ formula or term is closed or ground if it has no free variables. 
In our framework variables are only bound by $\now{x}{o}$, so in the following we
will consider a function $\mi{fv}(\varphi)$ that computes the free variables of $\varphi$, discarding the appearances of $x$ and $o$ from $\psi$ when
$\now{x}{o}\psi$ is found and collecting the rest of them, including those appearing in
temporal connectives.


We will only consider closed formulas in the following. Moreover, we will use
the notation $\varphi[b \mapsto v_1, r \mapsto v_2] \equiv
(\varphi[b \mapsto v_1])[r \mapsto v_2]$ to indicate that $b$ and $r$ are substituted by $v_1$ and $v_2$, respectively. A detailed explanation on how to compute the free variables and how to apply substitutions is available in~\cite{tplpTR}.

\paragraph{Logic for finite words} 
In order to evaluate our formulas, we need a way to interpret the timed terms that we use as the alphabet. In line with classical formulations 
of first order Boolean logic \cite{smullyan1995first}, formulas are evaluated in the context of an \emph{interpretation structure} $\ase$, which is 
a pair 
$(\sed, \sei)$ where $\sed$ is a non-empty set that is used as the interpreting domain, and $\sei$ is an 
interpretation function that assigns to each $f \in \fs^n$ an interpreting function $\sei(f) : \sed^n \rightarrow \sed$, and to each 
$p \in \pred^n$ an interpreting relation $\sei(p) \subseteq \sed^n$. These interpretations are naturally applied to closed terms by 
induction on the structure of terms as  $\den{f(e_1, \ldots, e_n)}^\ase = \sei(f)(\den{e_1}^\ase, \ldots, \den{e_1}^\ase) \in \sed$. 
Our logic 
proves \emph{judgments} of the form $\jud{u, i}{\varphi}{v}{\ase}$ that state that considering the finite word $u \in \letter^*$
from the position of its $i$-th letter, the formula $\varphi \in \form$ is evaluated to the truth value 
$v \in \{\true, \false, ?\}$ under the interpretation $\ase$. In other words, if we stand at the $i$-th 
letter of $u$ and start evaluating $\varphi$, moving forward in $u$ one letter at a time as time progresses, and using $\ase$ to interpret the 
terms that appear in the word and in the formula, we end up
getting the truth value $v$. Note that in our judgments the same interpretation structure holds ``eternally'' constant for all instants, while 
only one letter of $u$ is occurring at each instant. This is modeling what happens during the testing of a stream processing system: the code that 
defines how the program reacts to its inputs is the same during the execution of the program---which is modeled by a constant interpretation
structure---, while the inputs of the program and their corresponding output change with time ---which is modeled by the sequence of letters 
that is the word. This is not able to model updates in the program code, but it is expressive enough to be used during unit and integration testing, 
where the program code is fixed. Note the predicate symbols used in the formula correspond to the assertions used in the tests \cite{etorreborre2015Specs2}, 
whose meaning is also constant during the test execution. Judgments are defined by the following rules, where only the first rule that fits should be 
applied, and we assume $\ase = (\sed, \sei)$:

$$
\jud{u, i}{v}{v}{\ase} \text{~~~~if } v \in \{\false, \true\}
$$


$$
\jud{u, i}{p(e_1, \ldots, e_n)}{
\left\{\begin{array}{ll}
  \true & \text{~if~}(\den{e_1}^\ase, \ldots, \den{e_n}^\ase) \subseteq \sei(p)\\ 
  \false & \text{~otherwise}
\end{array}\right.
}{\ase} 
$$

$$
\jud{u, i}{e_1 = e_2}{
\left\{\begin{array}{ll}
  \true & \text{~if~} \den{e_1}^\ase = \den{e_2}^\ase\\ 
  \false & \text{~otherwise}
\end{array}\right.
}{\ase} 
$$

%


$$
\jud{u, i}{\!\!\!~\now{x}{o}.\varphi}{
\left\{\begin{array}{ll}
  v & \text{~if~} i \leq len(u) \wedge \jud{u, i+1}{\varphi[x \mapsto e_i, o \mapsto t_i]}{v}{\ase} \text{~~for~} u_i \equiv (e_i, t_i) \\ 
  ? & \text{~otherwise} \\
\end{array}\right.
}{\ase} 
$$


$$
\jud{u, i}{\next \varphi}{v}{\ase} \text{~~~~if } \jud{u, i+1}{\varphi}{v}{\ase}
$$


$$
\jud{u, i}{\teventually{t} \varphi}{
\left\{\begin{array}{ll}
\true & \text{ if } \exists k \in [i, i+(t-1)].~\jud{u, k}{\varphi}{\true}{\ase} \\
\false & \text{ if } \forall k \in [i, i+(t-1)].~\jud{u, k}{\varphi}{\false}{\ase}\\
? & \text{ otherwise } \\
\end{array}\right.
}{\ase} 
$$


$$
\jud{u, i}{\always{t} \varphi}{
\left\{
\begin{array}{ll}
\true & \text{ if } \forall k \in [i, i+(t-1)].~\jud{u, k}{\varphi}{\true}{\ase} \\
\false & \text{ if } \exists k \in [i, i+(t-1)].~\jud{u, k}{\varphi}{\false}{\ase} \\
? & \text{ otherwise } \\
\end{array}
\right.
}{\ase}
$$

%

$$
\jud{u, i}{\varphi_1 \tuntil{t} \varphi_2}{
\left\{
\begin{array}{ll}
\true & \text{ if } \exists k \in [i, i+(t-1)].~\jud{u, k}{\varphi_2}{\true}{\ase} ~\wedge \\
     & ~~~~~~~~\forall j \in [i, k).~ \jud{u, j}{\varphi_1}{\true}{\ase} \\
\false & \text{ if } \exists k \in [i, i+(t-1)].~\jud{u, k}{\varphi_1}{\false}{\ase} ~\wedge \\
     & ~~~~~~~~\forall j \in [i, k].~ \jud{u, j}{\varphi_2}{\false}{\ase} \\
\false & \text{ if } \forall k \in [i, i + (t-1)].~ \jud{u, k}{\varphi_1}{\true}{\ase} ~\wedge \\
     & ~~~~~~~~\forall l \in [i, min(i+(t-1), len(u))].~ \jud{u, l}{\varphi_2}{\false}{\ase} \\
? & \text{ otherwise } \\
\end{array}
\right.
}{\ase}
$$


$$
\jud{u, i}{\varphi_1 \trelease{t} \varphi_2}{
\left\{
\begin{array}{ll}
\true & \text{ if } \exists k \in [i, i+(t-1)].~\jud{u, k}{\varphi_1}{\true}{\ase} ~\wedge \\
     & ~~~~~~~~\forall j \in [i, k].~ \jud{u, j}{\varphi_2}{\true}{\ase}\\
\true & \text{ if } \forall k \in [i, i+(t-1)].~\jud{u, k}{\varphi_2}{\true}{\ase}\\
\false & \text{ if } \exists k \in [i, i+(t-1)].~\jud{u, k}{\varphi_2}{\false}{\ase} ~\wedge \\
     & ~~~~~~~~\forall j \in [i, k).~ \jud{u, j}{\varphi_1}{\false}{\ase} \\
? & \text{ otherwise } \\
\end{array}
\right.
}{\ase}
$$


We say $u \vDash^\ase \varphi$ iff $\jud{u,1}{\varphi}{\true}{\ase}$. 
%
The intuition underlying these definitions is that, if the word is too short to check
all the steps indicated by a temporal operator and neither $\true$ or $\false$ can be
obtained before finishing the word, then $?$ is obtained. Otherwise, the formula is
evaluated to
either $\true$ or $\false$ just by checking the appropriate sub-word. 
%
Note the consume operator ($\now{x}{o}$) is the only one that accesses the word directly, and that consume is equivalent to next applied to the corresponding formula at its body: for example $\jud{0~\epsilon, 1}{\now{x}{o}.x = 0}{v}{\ase} \iff \jud{0~\epsilon, 1}{\next (0 = 0)}{v}{\ase}$. 
It is trivial to check that timeless formulas---i.e.\ without temporal connectives---are always evaluated to one of the usual binary truth values $\true$ of $\false$, and that 
timeless formulas are evaluated to the same truth value irrespective of the word $u$ and the position
$i$ considered, even for $u \equiv \epsilon$ or $i > len(u)$. As a consequence, some temporal formulas are true even for words with a length smaller than the number of letters referred by the temporal connectives in the formula: for example, for any  $i$ and $\ase$ we have $\jud{\epsilon, i}{\next \true}{\true}{\ase}$---\emph{next} inspects the second letter, but the formula is true for the empty word because the body is trivially true---, $\jud{0~1~\epsilon, i}{\always{10}~(0 == 0)}{\true}{\ase}$---this \emph{always} refers to 10 letters, but it holds for a word with just 2 letters because the body is a tautology---, and similarly $\jud{0~\epsilon, 1}{\now{x}{o}.(x=0)}{\true}{\ase}$ because $\jud{0~\epsilon, 2}{0=0}{\true}{\ase}$.

The resulting logic gives some structure to letters and words, but it is not fully a first order logic because it does not provide neither existential or universal quantifiers for words. The \emph{consume} operator is somewhat a universal quantifier for letters, but can also be understood as a construct for parameter passing, like the binding operator from  TraceContract \cite{TraceContract}. 

%

Let us consider some example judgments for simple formulas, to start tasting the flavor of this logic. 

\begin{example}\label{ex:formulas}
Assume the set of constants $\{a, b, c\} \subseteq \fs^0$,
the set of variables $\{x,y,z,o,p,q\}$, and an interpretation structure 
$\mathcal{A} \equiv (\sed, \sei)$ for the initial model where $\sed = \fs^0$ and $\forall f \in \fs^0.~\sei(f) = f$. Then for the word $u \equiv \printarray[7ex]{{(b, 0)},{(b, 2)}}$
$\printarray[7ex]{{(a, 3)}}$ $\printarray[7ex]{{(a, 6)}}$ we can construct the following formulas:
\begin{itemize}
\item
$\juda{u}{\teventually{4}\ \now{x}{o}. x = c}{\false}$, since $c$ does not appear in the
first four letters.

\item
$\juda{u}{\teventually{5}\ \now{x}{o}. x = c}{?}$, since we have consumed the word,
$c$ did not appear in those letters but the timeout has not expired.


\item
$\juda{u}{\always{5}\ \now{x}{o}. (x = a \vee x = b)}{?}$, since the property holds until the
word is consumed, but the user required more steps.


\item
$\juda{u}{\now{x}{o}. x = b \tuntil{2} \now{y}{p}. y = a}{\false}$, 
since $a$ appears in the third letter, but the user wanted to check just the first two letters.

\item
$\juda{u}{\now{x}{o}. x = b \tuntil{5} \now{y}{p}. y = a}{\true}$, 
since $a$ appears in the third letter and, before that, $b$ appeared in all the letters.

\item
$\juda{u}{\now{x}{o}. x = a \trelease{2} \now{y}{p}. y = b}{\true}$,
since $b$ appears in all the required letters.


\item
$\juda{u}{\always{3}(\now{x}{o}.x = a) \rightarrow \next (\now{y}{p}. y = a)}{\true}$,
since the formula holds in the first three letters (note that the fourth letter
is required, since the formula involves the next operator).


\item
$\juda{u}{\always{2}(\now{x}{o}. x = b) \rightarrow (\teventually{2} \now{y}{p}. y = a)}{\false}$,
since in the first letter we have $b$ but we do not have $a$ until the third letter.

\item
$\juda{u}{(\now{x}{o}. x = b) \tuntil{2} \next (\now{y}{p}. y = a \wedge 
\next \now{z}{q}. z = a)}{\true}$,
since $\next (\now{y}{p}. y = a \wedge \next \now{z}{q}. z = a)$
holds in the second letter (that is, $(\now{y}{p}. y = a \wedge \next \now{z}{q}. z = a)$ 
holds in the third letter, which can be also understood as $a$ appears in the third and 
fourth letters).

\item
$\juda{u}{\now{x}{o}. \always{o + 6}~{x = b}}{\true}$, since the first letter is 
$b$ and hence the equality is evaluated to $\true$.





\end{itemize}

By using functions with arity greater than $0$, and predicate symbols, we can construct more complex formulas. For example given $\mathbb{N} \subseteq \fs^0, \mi{plus} \in \fs^2, \mi{leq} \in \pred^2$ and an interpretation structure $(\sed, \sei)$ where $\sed = \mathbb{N}$, $\forall n \in \mathbb{N}.~\sei(n) = n$, $\sei(\mi{plus})(x, y) = x + y$, $\sei(\mi{leq}) = \{(x, y) \in \mathbb{N} \times \mathbb{N} ~|~ x \leq y \}$, then we have 
$
\juda{\printarray[7ex]{{(0, 0)},{(1, 2)},{(2, 3)}}}{\teventually{2}\ \now{x}{o_1}.\now{y}{o_2}. leq(5, plus(x,y)}{\true}
$
\end{example}

For some examples in this paper we will assume \emph{the Spark interpretation structure $\sse$}, that captures the observable semantics of a Spark program, and where timestamps are interpreted as Unix timestamps as usual in Java.
We will not provide a formalization of $\sse$, but the idea is that the prototype 
we present in Section \ref{sec:tool} is intended to implement a procedure to prove judgments under the Spark interpretation structure. This interpretation uses the set of Scala expressions as the domain $\sed$, and assumes that letters are timed tuples of terms, and that each input or output DStream has an assigned tuple index, so that each element of the tuple represents the micro batch at that instant for the corresponding DStream. This 
is expressive enough to express any Spark Streaming program, because the set of DStreams is fixed during the lifetime of a Spark Streaming application. 
Let us see some simple formulas we can write with this logic and in our prototype. 

\begin{example}\label{ex:simpleFormulasPF}
Assuming a Spark Streaming program with one input DStream and one output DStream, the formula $\varphi_1$ below expresses the requirement that the output DStream will always contain numbers greater 
than 0, for 10 batches. As we have one input and one output, $\sse$ uses
timed  letters in 
$\mathbb{U} \times \mathbb{N}$ for $\mathbb{U} = \mathbb{RDD} \times \mathbb{RDD}$ and  $\mathbb{RDD}$ the set of all Spark RDDs.
$$
\begin{array}{l}
\sei(\mathit{allOutValuesGtZero}) = \{ (rdd_i, rdd_o) \in \mathbb{U} ~|~ \forall e \in rdd_o.~ e \in \mathbb{Z} \wedge e > 0 \} \\
\varphi_1 = \always{10}~\now{\mathit{rdds}}{o}.{\it allOutValuesGtZero}(\mathit{rdds})
\end{array}
$$

\noindent This formula can be written in our prototype as follows: 

{\codesize
\begin{verbatim}
always(nowTime[(RDD[Int], RDD[Int])]{ (letter, time) => 
  letter._2 should foreachRecord { _ > 0} 
}) during 10
\end{verbatim}
}

The formula $\varphi_2$ below 
expresses that time always increases monotonically during 10 instants: 
$$
\begin{array}{l}
\sei(\mathit{leq}) = \{ (o_1, o_2) \in \mathbb{N} \times \mathbb{N} ~|~ o_1 \leq o_2 \} \\
\varphi_2 = \always{9}~\now{x_1}{o_1}.\now{x_2}{o_2}.\mathit{leq}(o_1, o_2)
\end{array}
$$

\noindent which we can express in our prototype as:
{\codesize
\begin{verbatim}
always(nextTime[(RDD[Int], RDD[Int])]{ (letter, time) => 
  nowTime[U]{ (nextLetter, nextTime) =>
    time.millis <= nextTime.millis
  } 
}) during 9
\end{verbatim}
}
\end{example}

Once the formal definition has been presented, we require a decision procedure for
evaluating formulas. 
Next we present a formula evaluation algorithm inferred from
the logic presented above. 

\paragraph{Decision procedure for $\sstl$} 
Just like ScalaCheck \cite{nilsson2014scalacheck} and any other testing tool of the QuickCheck
family~\cite{claessen2011quickcheck,papadakis2011proper}, this decision procedure does not try to
be complete for proving the veritative value of formulae, but just to be complete for failures, i.e.\ judgments to the truth value $\false$. 
%
For this purpose we 
define an abstract rewriting system for reductions
$u \vDash^\ase \varphi \leadsto^* v$ for $v$ in the same domain as above. We write $u \vDash \varphi \leadsto^* v$ when the interpretation $\ase$ is implied by the context. 
Given a letter $a \in \Sigma$, a word $u \in \Sigma^*$, 
a set of terms $e,e_1,\ldots,e_n \in \term$,
a timeout $t \in \mathbb{N}^+$,
and formulas $\varphi, \varphi_1, \varphi_2 \in \sstl$,
we have the following rules:\footnote{Formulas built with propositional operators just
evaluate the sub-formulas and apply the connectives as usual.}

\begin{enumerate}
\item
Rules for $u \vDash^{\ase} p(e_i)$:

$
\begin{array}{lllll}
1) & u \vDash^{\ase} p(e_1,\ldots,e_n) & \leadsto & \true & 
\text{if~}(\den{e_1}^\ase, \ldots, \den{e_n}^\ase) \subseteq \sei(p)\\
2) & u \vDash^{\ase} p(e_1,\ldots,e_n) & \leadsto & \false & \textrm{otherwise}
\end{array}
$

\item
Rules for $u \vDash e_1 = e_2$:

$
\begin{array}{llll}
1) & u \vDash^{\ase} e_1 = e_2 & \leadsto & \den{e_1}^\ase = \den{e_2}^\ase\\
\end{array}
$

\item
Rules for $u \vDash \now{x}{o}.\varphi$:

$
\begin{array}{llll}
1) & \epsilon \vDash \now{x}{o}.\varphi & \leadsto & ?\\
2) & (e, t) u \vDash \now{x}{o}. \varphi & \leadsto & 
        u \vDash \varphi[x \mapsto e][o \mapsto t] \\
\end{array}
$
%
%

\item
Rules for $u \vDash \next\ \varphi$:

$
\begin{array}{llll}
1) & \epsilon \vDash \next\ \varphi & \leadsto & \epsilon \vDash \varphi\\
2) & a u \vDash \next\ \varphi & \leadsto & u \vDash \varphi \\
\end{array}
$

\item
Rules for $u \vDash \teventually{t}\ \varphi$:

$
\begin{array}{lllll}
1) & \epsilon \vDash \teventually{t}\ \varphi & \leadsto & \epsilon \vDash \varphi &\\
2) & u \vDash \teventually{0}\ \varphi & \leadsto & \false & \\
3) & u \vDash \teventually{t}\ \varphi & \leadsto & \true & 
\text{if } u \vDash \varphi \leadsto^* \true  \\
4) & a u \vDash \teventually{t}\ \varphi & \leadsto & u \vDash \teventually{t - 1}\ \varphi &\text{if } a u \vDash \varphi \leadsto^* \false  \\
\end{array}
$

\item
Rules for $u \vDash \always{t}\ \varphi$:

$
\begin{array}{lllll}
1) & \epsilon \vDash \always{t}\ \varphi & \leadsto & \epsilon \vDash \varphi &\\
2) & u \vDash \always{0}\ \varphi & \leadsto & \true &\\
3) & u \vDash \always{t}\ \varphi & \leadsto & \false &
\text{if } u \vDash \varphi \leadsto^* \false \\
4) & a u \vDash \always{t}\ \varphi & \leadsto & u \vDash \always{t-1}\ \varphi &
\text{if } a u \vDash \varphi \leadsto^* \true  \\
\end{array}
$

\item
Rules for $u \vDash \varphi_1 \tuntil{t} \varphi_2$:

$
\begin{array}{lllll}
1) & \epsilon \vDash \varphi_1 \tuntil{t} \varphi_2 & \leadsto & \epsilon \vDash \varphi_2 &\\
2) & u \vDash \varphi_1 \tuntil{0} \varphi_2 & \leadsto & \false & \\
3) & u \vDash \varphi_1 \tuntil{t} \varphi_2 & \leadsto & \true &
\text{if } u \vDash \varphi_2 \leadsto^* \true \\
4) & u \vDash \varphi_1 \tuntil{t} \varphi_2 & \leadsto & \false &
\text{if } u \vDash \varphi_1 \leadsto^* \false \wedge u \vDash \varphi_2 \leadsto^* \false \\
5) & a u \vDash \varphi_1 \tuntil{t} \varphi_2 & \leadsto & u \vDash 
\varphi_1 \tuntil{t-1} \varphi_2 &
\text{if } a u \vDash \varphi_1 \leadsto^* \true \wedge a u \vDash \varphi_2 \leadsto^* \false\\
\end{array}
$

\item
Rules for $u \vDash \varphi_1 \trelease{t} \varphi_2$:

$
\begin{array}{lllll}
1) & \epsilon \vDash \varphi_1 \trelease{t} \varphi_2 & \leadsto & \epsilon \vDash \varphi_1 &\\
2) & u \vDash \varphi_1 \trelease{0} \varphi_2 & \leadsto & \true \\
3) & u \vDash \varphi_1 \trelease{t} \varphi_2 & \leadsto & \true &
\text{if } u \vDash \varphi_1 \leadsto^* \true \wedge u \vDash \varphi_2 \leadsto^* \true \\
4) & u \vDash \varphi_1 \trelease{t} \varphi_2 & \leadsto & \false &
\text{if } u \vDash \varphi_2 \leadsto^* \false \\
5) & a u \vDash \varphi_1 \trelease{t} \varphi_2 & \leadsto & u \vDash 
\varphi_1 \trelease{t-1} \varphi_2 &
\text{if } a u \vDash \varphi_1 \leadsto^* \false \wedge a u \vDash \varphi_2 \leadsto^* \true \\
\end{array}
$

\end{enumerate}

\noindent
for $\epsilon$ the empty word. These rules follow this schema: (i) an inconclusive value is returned when
the empty word is found; (ii) the formula is appropriately evaluated when the timeout expires; (iii) it
evaluates the subformulas to check whether a value can be obtained; it consumes the current letter and
continues the evaluation; and (iv) inconclusive is returned if the subformulas are evaluated to inconclusive
as well, and hence the previous rules cannot be applied.
%
%
Hence, note that these rules have conditions that depend on the future. This happens in rules with a
condition involving $\leadsto^*$ that inspects not only the first letter of the word, i.e., what is happening now, but also the subsequent letters, as illustrated by the following examples:

\begin{example}\label{ex:eval1}
We recall the word $u \equiv \printarray[7ex]{{(b, 0)},{(b, 2)},{(a, 3)},{(a, 6)}}$
from Example~\ref{ex:formulas} and evaluate the following formulas:
\begin{itemize}
\item
$\printarray[7ex]{{(b, 0)},{(b, 2)},{(a, 3)},{(a, 6)}}
\vDash \always{2}(\now{x}{o}. x = b) \rightarrow (\teventually{2}~\now{y}{p}. y = a) \leadsto \false$, because first the $x$ in $(\now{x}{o}. x = b)$
is bound to $b$ and hence the premise holds, but 
$\printarray[7ex]{{(b, 0)},{(b, 2)},{(a, 3)},{(a, 6)}} \vDash 
(\teventually{2} \now{y}{p}. y = a) \leadsto\\
\printarray[7ex]{{(b, 2)},{(a, 3)},{(a, 6)}} \vDash (\teventually{1}~\now{y}{p}. y = a)
\leadsto\\
\printarray[7ex]{{(a, 3)},{(a, 6)}} \vDash
(\teventually{0} \now{y}{p}. y = a) \leadsto \false$.

\item
$\printarray[7ex]{{(b, 0)},{(b, 2)},{(a, 3)},{(a, 6)}} \vDash 
(\now{x}{o}. x = b) \tuntil{2} \next (\now{y}{p}. y = a \wedge \next \now{z}{q}. z = a)
\leadsto \\
\printarray[7ex]{{(b, 2)},{(a, 3)},{(a, 6)}} \vDash 
(\now{x}{o}. x = b) \tuntil{1} \next (\now{y}{p}. y = a \wedge \next \now{z}{q}. z = a)$,
which requires to check the second and third letters to check that the second formula does
not hold. Then we have 
$\printarray[7ex]{{(b, 2)},{(a, 3)},{(a, 6)}} \vDash 
(\now{x}{o}. x = b) \tuntil{1} \next (\now{y}{p}. y = a \wedge \next \now{z}{q}. z = a)
\leadsto
\true$ after checking the third and fourth letters.

\item
$\printarray[7ex]{{(b, 0)},{(b, 2)},{(a, 3)},{(a, 6)}} \vDash 
\now{x}{o}. \always{o + 6}{x = b}
\leadsto\\
\printarray[7ex]{{(b, 2)},{(a, 3)},{(a, 6)}} \vDash
\always{6}{\true}$, just by binding the variables.
Then we have $\printarray[7ex]{{(b, 2)},{(a, 3)},{(a, 6)}} \vDash
\always{6}{\true}
\leadsto\\
\printarray[7ex]{{(a, 3)},{(a, 6)}} \vDash \always{5}{\true}
\leadsto\\
\printarray[7ex]{{(a, 6)}} \vDash \always{4}{\true}
\leadsto
\varepsilon \vDash \always{3}{\true}
\leadsto
\varepsilon \vDash \true
\leadsto
\true
$ just by applying the rules for $\square$.

\end{itemize}
\end{example}

To use this procedure as the basis for our implementation, we would had to keep a list of suspended alternatives for each of the rules above, that are pending the resolution of the conditions that define each alternative, which will be solved in the future. For example if we apply rule 5 to an application of $\teventually{t}$ for a non empty word and $t >0$ then we get 2 alternatives for sub-rules 5.3 and 5.4, and those alternatives will depend on whether the nested formula $\varphi$ is reduced to $\top$ or $\perp$ using $\leadsto^*$, which cannot be determined at the instant when rule 5 is applied. 
This is because, although we do have all the batches for a generated test case corresponding to an input stream, the batches for output streams generated
by transforming the input will be only generated after waiting the corresponding number of instants, 
as our implementation runs the actual code that is the subject of the test in a 
 local Spark cluster. 
This leads to a complex and potentially expensive computation, since many pending possible alternatives have to be kept open. 
Instead of using this approach,
it would be much more convenient to define a \emph{stepwise} method 
with transition rules that only inspect the first letter of the input word. 

\subsection{A transformation for stepwise evaluation\label{subsec:stepwise}}

In order to define this stepwise evaluation, it is worth noting that all the properties
are finite (that is, all of them can be proved or disproved after a finite number of steps).
It is hence possible to express any formula only using the temporal operators 
next and consume, which leads us
to the following definition.

\begin{definition}[Next form]
We say that a formula $\psi \in \sstl$ is in \emph{next form} iff.\ it is built by using the
following grammar:
$$
\begin{array}{ll}
\psi ::= & \false \mid \true \mid p(e, \ldots, e) \mid e = e \mid \neg \psi \mid \psi \vee \psi
\mid \psi \wedge \psi \mid \psi \rightarrow \psi \mid \next \psi \mid \now{x}{o}.\psi\\
\end{array}
$$
\end{definition}

We can extend the transformation in \cite{sscheckIFM}
for computing the next form of any formula $\varphi \in \sstl$:

\begin{definition}[Explicit next transformation]
Given 
a formula $\varphi \in \sstl$, the function $\nt^e(\varphi)$ computes another formula
$\varphi' \in \sstl$, such that $\varphi'$ is in \emph{next form} and
$\forall u \in \Sigma^*. u \vDash \varphi \iff u \vDash \varphi'$.

$
\begin{array}{llll}
\nt^e(\varphi) & = & \varphi & \textrm{ if } \varphi \in \{\true, \false,
p(e_1, \ldots, e_n), e_1 = e_2\}\\
\nt^e(\mathit{op}\ \varphi) & = & \mathit{op}\ \nt^e(\varphi) & \textrm{ if } 
\mathit{op} \in \{\neg, \next, \now{x}{o}\}\\
\nt^e(\varphi_1\ \mathit{op}\ \varphi_2) & = & \nt^e(\varphi_1)\ \mathit{op}\
\nt^e(\varphi_2) & \textrm{ if } \mathit{op} \in \{\vee, \wedge, \rightarrow\}\\
\end{array}
$

$
\begin{array}{lll}
\nt^e(\teventually{t} \varphi) & = &
          \nt^e(\varphi) \vee \next \nt^e(\varphi) \vee \ldots \vee 
          \next^{t-1} \nt^e(\varphi)\\
\nt^e(\always{t} \varphi) & = &
          \nt^e(\varphi) \wedge \next \nt^e(\varphi) \wedge \ldots \wedge 
          \next^{t-1} \nt^e(\varphi)\\
\nt^e(\varphi_1 \tuntil{t} \varphi_2) & = &
          \nt^e(\varphi_2) \vee (\nt^e(\varphi_1) \wedge \next \nt^e(\varphi_2)) \vee\\
          &&
          (\nt^e(\varphi_1) \wedge \next \nt^e(\varphi_1) \wedge \next^{2} \nt^e(\varphi_2))
          \vee \ldots \vee\\
          &&
          (\nt^e(\varphi_1) \wedge \next \nt^e(\varphi_1) \wedge \ldots \wedge
          \next^{t-2} \nt^e(\varphi_1) \wedge \next^{t-1} \nt^e(\varphi_2))\\
\nt^e(\varphi_1 \trelease{t} \varphi_2) & = &
          (\nt^e(\varphi_2) \wedge \next \nt^e(\varphi_2) \wedge \ldots \wedge \next^{t-1} \nt^e(\varphi_2)) \vee\\
          &&
          (\nt^e(\varphi_1) \wedge \nt^e(\varphi_2)) \vee
          (\nt^e(\varphi_2) \wedge \next (\nt^e(\varphi_1) \wedge \nt^e(\varphi_2))) \vee\\
          &&
          (\nt^e(\varphi_2) \wedge \next \nt^e(\varphi_2) \wedge \next^{2} (\nt^e(\varphi_1) \wedge \nt^e(\varphi_2)))
          \vee \ldots \vee\\
          &&
          (\nt^e(\varphi_2) \wedge \next \nt^e(\varphi_2) \wedge \ldots \wedge
          \next^{t-2} \nt^e(\varphi_2) \wedge \next^{t-1} (\nt^e(\varphi_1) \wedge \nt^e(\varphi_2))\\
\end{array}
$
\noindent
for $e_1, e_2 \in  \term$, $x, o \in \mathcal{V}$, $p \in \pred^n$, and
$\varphi, \varphi_1, \varphi_2 \in \sstl$.
\end{definition}

Note that (i) it is not always possible to compute the next form a priori, since the
time in temporal operators might contain variables that need to be bound
and (ii) the transformation might produce large formulas. 
For these reasons, it is worth transforming the
formula following a lazy strategy, which only generates the subformulas required in the
current and the next states.
We present next a recursive function
that allows us to compute the next form in a lazy way, which we use to improve the efficiency of 
our prototype, as we will see in Section~\ref{subsec:design}.

\begin{definition}[Recursive next transformation]\label{def:recNextForm}
Given 
a formula $\varphi \in \sstl$, the function $\nt(\varphi)$ computes another formula
$\varphi' \in \sstl$, such that $\varphi'$ is in \emph{next form} and
$\forall u \in \Sigma^*. u \vDash \varphi \iff u \vDash \varphi'$.
$$
\begin{array}{llll}
\nt(\varphi) & = & \varphi & \textrm{if } \varphi \in \{\true, \false,
p(e_1, \ldots, e_n), e_1 = e_2\}\\
\nt(\mathit{op}\ \varphi) & = & \mathit{op}\ \nt(\varphi) & \textrm{if } 
\mathit{op} \in \{\neg, \next, \now{x}{o}\}\\
\nt(\varphi_1\ \mathit{op}\ \varphi_2) & = & \nt(\varphi_1)\ \mathit{op}\
\nt(\varphi_2) & \textrm{if } \mathit{op} \in \{\vee, \wedge, \rightarrow\}\\
\nt(\teventually{1} \varphi) & = &
          \nt(\varphi) &\\
\nt(\teventually{t} \varphi) & = &
          \nt(\varphi) \vee \next \nt(\teventually{t - 1} \varphi) & \textrm{if $t \geq 2$}\\
\nt(\always{1} \varphi) & = &
          \nt(\varphi) &\\
\nt(\always{t} \varphi) & = &
          \nt(\varphi) \wedge \next \nt(\always{t - 1} \varphi) & \textrm{if $t \geq 2$}\\
\nt(\varphi_1 \tuntil{1} \varphi_2) & = & \nt(\varphi_2) & \\
\nt(\varphi_1 \tuntil{t} \varphi_2) & = &
          \nt(\varphi_2)\ \vee\\
          & & (\nt(\varphi_1) \wedge \next \nt(\varphi_1 \tuntil{t-1} \varphi_2))
          & \textrm{if $t \geq 2$}\\
\nt(\varphi_1 \trelease{1} \varphi_2) & = &
          \nt(\varphi_1) \wedge \nt(\varphi_2) &\\
\nt(\varphi_1 \trelease{t} \varphi_2) & = &
          (\nt(\varphi_1) \wedge \nt(\varphi_2)) \vee\\
          && (\nt(\varphi_2) \wedge \next \nt(\varphi_1 \trelease{t-1} \varphi_2)) &
          \textrm{if $t \geq 2$}\\
\end{array}
$$
\noindent
for $e_1, \ldots e_n \in  \term$, $x, o \in \mathcal{V}$, $p \in \pred^n$, and
$\varphi, \varphi_1, \varphi_2 \in \sstl$.
\end{definition}


Next, we present some results about these transformations and an auxiliary lemma
that indicates that, if two formulas are equivalent at time $1$, then
they keep being equivalent the rest of the execution:

\begin{lemma}\label{lemma:n}
Given an alphabet $\Sigma$ and
formulas $\varphi, \varphi' \in \sstl$,
if $\forall u \in \Sigma^*. u,1 \vDash \varphi \iff u,1 \vDash \varphi'$
then $\forall u \in \Sigma^*, \forall n \in \mathbb{N}^+. u,n \vDash \varphi \iff u,n \vDash \varphi'$.
\end{lemma}

\begin{theorem}[Transformation equivalence]\label{th:t_eq}
Given a formula $\varphi \in \sstl$ such that $\varphi$ does not contain variables in temporal
connectives, we have $\nt(\varphi) = \nt^e(\varphi)$.
\end{theorem}

It is straightforward to see that the formula obtained by this transformation is in
\emph{next form},
since it only introduces formulas using the 
temporal operators next or consume. The equivalence between formulas is
stated in Theorem~\ref{th:equiv_next}:

\begin{theorem}\label{th:equiv_next}

Given an alphabet $\Sigma$,
an interpretation $\ase$, and
formulas $\varphi, \varphi' \in \sstl$, such that $\varphi' \equiv \nt(\varphi)$,
we have 
$\forall u \in (\Sigma \times \mathbb{N})^*. u \vDash^{\ase} \varphi \iff u \vDash^{\ase} \varphi'$.
\end{theorem}

\noindent
Both theorems are proved by induction in the structure of formulas and using 
Lemma~\ref{lemma:n}. 
Detailed proofs are available in~\ref{app:proofs}.

The show next some examples of explicit transformation and the first step of the lazy transformation.

\begin{example}\label{ex:nt}
We present here how to transform some of the formulas from Example~\ref{ex:formulas}.
Note that the last one cannot be completely transformed a priori:
\begin{itemize}
\item
$\nt^e(\teventually{4}\ \now{x}{o}. x = c) = 
\now{x}{o}. x = c \vee
\next \now{x}{o}. x = c \vee
\next^2 \now{x}{o}. x = c \vee 
\next^3 \now{x}{o}. x = c$


\item
$\nt^e(\now{x}{o}. x = b \tuntil{2} \now{x}{o}. x = a) =
\now{x}{o}. x = a \vee (\now{x}{o}. x = b \wedge \next \now{x}{o}. x = a)$



\item
$\nt^e(\always{2}(\now{x}{o}. x = b) \rightarrow (\teventually{2} \now{y}{p}. y = a) =
(\now{x}{o}. x = b \rightarrow (\now{y}{p}. y = a \vee \next \now{y}{p}. y = a)) \wedge
\next (\now{x}{o}. x = b \rightarrow (\now{y}{p}. y = a \vee \next \now{y}{p}. y = a))$

\item
$\nt^e((\now{x}{o}. x = b) \tuntil{2} \next (\now{y}{p}. y = a \wedge 
\next \now{z}{q}. z = a)) = 
\next (\now{y}{p}. y = a \wedge \next \now{z}{q}. z = a) \vee
(\now{x}{o}. x = b \wedge \next^2 (\now{y}{p}. y = a \wedge \next \now{z}{q}. z = a))$
                 
\end{itemize}
\end{example}


\begin{example}\label{ex:lazy_nt}
We present the lazy next transformation for some formulas, where we just apply the first
transformation. Note that in the last example it is not possible to compute the next
form in an eager way:
\begin{itemize}

\item
$\nt(\always{2}(\now{x}{o}. x = b) \rightarrow (\teventually{2} \now{y}{p}. y = a))
=
(\now{x}{o}. x = b) \rightarrow (\teventually{2} \now{y}{p}. y = a) \wedge
\next \nt(\always{1}(\now{x}{o}. x = b) \rightarrow (\teventually{2} \now{y}{p}. y = a))
$

\item
$\nt(\now{x}{o}. \always{o + 6}{x = b}) = 
\now{x}{o}. \nt(\always{o + 6}{x = b})
$
\end{itemize}
\end{example}

Once the next form of a formula has been computed, it is possible to evaluate it for a
given word just by traversing its letters. We just evaluate the atomic formulas in the
present moment (that is, those properties that does not contain the next operator) and
remove the next operator otherwise, so these properties will be evaluated for the next
letter. This method is detailed as follows:

\begin{definition}[Letter simplification]\label{def:let_simp}
Given a formula $\psi$ in next form, a letter $s \in \Sigma$,
where $s$ can be either $(e, t)$, with $e \in \term, t \in \mathbb{N}$,
or the empty letter $\emptyset$,
and an interpretation $\ase = (\sed, \sei)$, the function $\mathit{ls}^\ase(\psi, s)$
($\mathit{ls}(\psi, s)$ when $\ase$ is clear from the context)
\emph{simplifies $\psi$ with s} as follows:

$
\begin{array}{llll}
\mathit{ls}(\varphi, s) & = & \varphi & \textrm{if } \varphi \in \{\true, \false\}\\

%

\mathit{ls}^\ase(p(e_1, \ldots, e_n), s) & = &
 (\den{e_1}^\ase, \ldots, \den{e_n}^\ase) \subseteq \sei(p)&\\


\mathit{ls}^\ase(e_1 = e_2, s) & = & \mathit{synEq}(\den{e_1}^\ase, \den{e_2}^\ase) &\\


%
%

\mathit{ls}(\psi_1\ \mathit{op}\ \psi_2, s) & = & \mathit{ls}(\psi_1, s)\ \mathit{op}\ \mathit{ls}(\psi_2, s)
& \textrm{if } \mathit{op} \in \{\vee, \wedge, \rightarrow\}\\

\mathit{ls}(\next \psi, (e, t)) & = & \psi &\\

\mathit{ls}(\next \psi, \emptyset) & = & \mathit{ls}(\psi, \emptyset)&\\


\mathit{ls}(\now{x}{o}. \psi, (e, t)) & = & \psi[x \mapsto e][o \mapsto t]&\\

\mathit{ls}(\now{x}{o}. \psi, \emptyset) & = & ? &\\
\end{array}
$
\end{definition}

\noindent
where $\mathit{synEq}$ stands for syntactic equality.
Note that using the empty letter forces the complete evaluation of the formula.
Using this function and applying propositional logic and the interpretation $\ase$ when
definite values are found it is possible to evaluate formulas in a step-by-step
fashion.\footnote{Note that the value ? is
only reached when the word is consumed and this simplification cannot be applied.}
In this way, we can solve the formulas from the previous example as illustrated in the next
example.

\begin{example}
We present here the lazy evaluation process for some formulas in Example~\ref{ex:eval1}
using the word $u \equiv 
\printarray[7ex]{{(b, 0)},{(b, 2)},{(a, 3)},{(a, 6)}}$.

\begin{itemize}
\item
$\nt(\always{2}(\now{x}{o}. x = b) \rightarrow (\teventually{2} \now{y}{p}. y = a))
=
(\now{x}{o}. x = b) \rightarrow (\now{y}{p}. y = a \vee 
      (\next\nt(\teventually{1} \now{y}{p}. y = a))) \wedge
\next \nt(\always{1}(\now{x}{o}. x = b) \rightarrow (\teventually{2} \now{y}{p}. y = a))
$ (from Example~\ref{ex:lazy_nt}).
\begin{itemize}
\item
$\mathit{ls}((\now{x}{o}. x = b) \rightarrow (\now{y}{p}. y = a \vee 
      (\next\nt(\teventually{1} \now{y}{p}. y = a))) \wedge
\next \nt(\always{1}(\now{x}{o}. x = b) \rightarrow (\teventually{2} \now{y}{p}. y = a)),
     (b, 0))) = \textrm{(consume letter)}\\
     \true \rightarrow (\false \vee 
     \nt(\teventually{1} \now{y}{p}. y = a)) \wedge
     \nt(\always{1}(\now{x}{o}. x = b) \rightarrow (\teventually{2} \now{y}{p}. y = a))
     \\\equiv \textrm{(simplification)}\\
     \nt(\teventually{1} \now{y}{p}. y = a) \wedge
     \nt(\always{1}(\now{x}{o}. x = b) \rightarrow (\teventually{2} \now{y}{p}. y = a))
     \\ \equiv \textrm{(lazy evaluation of the next transformation)}\\
     \now{y}{p}. y = a \wedge
     \now{x}{o}. x = b \rightarrow (\now{y}{p}. y = a \vee 
     \next\nt(\teventually{1} \now{y}{p}. y = a)))$.

\item
$\mathit{ls}(\now{y}{p}. y = a \wedge
     \now{x}{o}. x = b \rightarrow (\now{y}{p}. y = a \vee 
     \next\nt(\teventually{1} \now{y}{p}. y = a))), (b, 2)) 
     \\= \textrm{(consume letter)}\\
     \false \wedge
     \true \rightarrow (\false \vee 
     \nt(\teventually{1} \now{y}{p}. y = a)))
     \\\equiv \textrm{(simplification)}\\
     \false$.
\end{itemize}
%
%

\item
$\nt(\now{x}{o}. \always{o + 6}{x = b}) = 
\now{x}{o}. \nt(\always{o + 6}{x = b})$ (from Example~\ref{ex:lazy_nt}).
\begin{itemize}
\item
$\mathit{ls}(\now{x}{o}. \nt(\always{o + 6}{x = b}), (b, 0)) = \nt(\always{6}{b = b})
     \\ \equiv \textrm{(lazy evaluation of the next transformation)}\\
     \true \wedge X \nt(\always{5}{b = b})
     \\\equiv \textrm{(simplification)}\\ X \nt(\always{5}{b = b})$

\item
$\mathit{ls}(X \nt(\always{5}{b = b}), (b, 2)) = \nt(\always{5}{b = b})
     \\ \equiv \textrm{(lazy evaluation of the next transformation and simplification)}\\
     X \nt(\always{4}{b = b})$

\item
$\mathit{ls}(X \nt(\always{4}{b = b}), (a, 3)) = \nt(\always{4}{b = b})
     \\ \equiv \textrm{(lazy evaluation of the next transformation and simplification)}\\
     X \nt(\always{3}{b = b})$

\item
$\mathit{ls}(X \nt(\always{3}{b = b}), (a, 6)) = \nt(\always{3}{b = b})
     \\ \equiv \textrm{(lazy evaluation of the next transformation and simplification)}\\
     X \nt(\always{2}{b = b})$

\item
$\mathit{ls}(X \nt(\always{3}{b = b}), \emptyset) = \nt(\always{3}{b = b})
     \\ \equiv \textrm{(lazy evaluation of the next transformation and simplification)}\\
     X \nt(\always{2}{b = b})$

\item
$\mathit{ls}(X \nt(\always{2}{b = b}), \emptyset) = 
\mathit{ls}(\nt(\always{2}{b = b}), \emptyset)
\\ \equiv \textrm{(lazy evaluation of the next transformation and simplification)}\\
\mathit{ls}(\nt(\always{1}{b = b}), \emptyset)
\\ \equiv \textrm{(lazy evaluation of the next transformation and simplification)}\\
\mathit{ls}(\true, \emptyset) \equiv \true
$

\end{itemize}
\end{itemize}
\end{example}


When no variables appear in the timeouts of temporal operators, 
the next transformation gives also the intuition that inconclusive values can be avoided
if we use a word as long as the number of next/consume operators nested in the
transformation.\footnote{Note that it might be possible to avoid an inconclusive value with shorter words, so this
is a \emph{sufficient} condition.} We define this \emph{safe word length} as follows:

\begin{definition}[Safe word length]\label{def:safe-word-length}
Given a formula $\varphi \in \sstl$ without variables in any timeouts of the temporal operators that occur in it, its longest
required check $\mathit{swl}(\varphi) \in \mathbb{N}$ is the maximum word length of a word
$u$ such that we have $u \vDash \varphi \in \{\true, \false\}$.
It is defined as follows:

$
\begin{array}{llll}
\mathit{swl}(\varphi) & = & 0 & \textrm{if } \varphi \in \{\true, \false,
p(e_1, \ldots, e_n), \\
&&& e_1 = e_2\}\\
\mathit{swl}(\neg \varphi) & = & \mathit{swl}(\varphi) &\\
\mathit{swl}(\varphi_1\ \mathit{op}\ \varphi_2) & = &
          \mathit{max}(\mathit{swl}(\varphi_1), \mathit{swl}(\varphi_2))    
        & \textrm{if } \mathit{op} \in \{\vee, \wedge, \rightarrow\}\\ 
\mathit{swl}(\mi{op}\ \varphi) & = &
          \mathit{swl}(\varphi) + 1 & 
          \textrm{if } \mathit{op} \in \{\next, \now{x}{o}\}\\
\mathit{swl}(\mi{op}_{t}\ \varphi) & = &
          \mathit{swl}(\varphi) + (t - 1) &
          \textrm{if } \mathit{op} \in \{\lozenge, \square\}\\
\mathit{swl}(\varphi_1 \tuntil{t} \varphi_2) & = &
          \mathit{max}(\mathit{swl}(\varphi_1), \mathit{swl}(\varphi_2)) + (t - 1) &
         \textrm{if } \mathit{op} \in \{U, R\} \\
\end{array}
$
\end{definition}

\begin{example}
We present here the safe word length for some of the formulas in Example~\ref{ex:formulas}:
\begin{itemize}









\item
$\mathit{swl}(\always{3}(\now{x}{o}.x = a) \rightarrow \next (\now{y}{p}. y = a)) = 4$.



\item
$\mathit{swl}((\now{x}{o}. x = b) \tuntil{2} \next (\now{y}{p}. y = a \wedge 
\next \now{z}{q}. z = a)) = 4$.
\end{itemize}
\end{example}

On the other hand, we cannot define a safe word length for arbitrary formulas with variables in timeouts, because an application of the consume operator might bind those variables using a letter of the input word, so there is no way to determine the value of the timeout for all possible words. 

\subsection{Generating words\label{subsec:generating_words}}

Besides stating properties, formulas can be used to generate
words.
%
%
%
In particular, we will generate sequences of terms from formulas; these sequences can
then be extended by pairing each letter
with a number generated by an arbitrary monotonically
increasing function, hence 
obtaining words with timed terms as letters. 
The formulas used for generating terms have the following restrictions:
\begin{itemize}
\item
Given a formula $\lambda_x^o. \varphi$, we have $o \not\in \mathit{fv}(\varphi)$.
Since in this stage we do not generate times, they cannot be used.

\item
Formulas do not contain the negation operator or the \texttt{false} constant. The process tries to generate words that make the 
formula evaluate to true, so we would not generate any word for a contradiction. Besides, we 
do not support negation because that would imply maintaining a set of excluded words, and we 
wanted to define simple ScalaCheck generators in a straightforward way.
\end{itemize}
For describing how the generators compute the sequences we first need to introduce a
constant $\mi{err} \in \term^*$ that stands for an erroneous sequence. Moreover, we use
the
notation $+ : \term^* \times \term^* \rightarrow \term^*$ ($u + \mi{err} = \mi{err} + u =
\mi{err}$) for composing words, and extend the union on $\term^*$ as:

$$
\begin{array}{lll}
a\ u \cup b\ v & = & (a \cup b) + u \cup v\\
u \cup \epsilon & = & u\\
u \cup \mi{err} & = & \mi{err}\\
\end{array}
$$

\noindent
for $a, b \in \term$ and $u, v \in \term^*$. Note that we assume 
that syntax for sets and unions is defined in $\fs$.
Using these ideas, we have:

 
\begin{definition}[Random word generation]\label{def:randWordGen}
Given an interpretation $\ase$,
$e_1, \ldots, e_n \in \term$,
$p \in \pred^n$,
formulas $\psi$, $\psi_1$, and
$\psi_2$ in next form, the function
$\mathit{gen}^\ase$ ($\mathit{gen}$ when $\ase$ is clear from the
context) generates a finite word $u \in \term^*$.
If different equations can
be applied for a given formula any of them can be chosen:

$
\begin{array}{llll}
\mathit{gen}(\true) & = & \emptyset &\\
\mathit{gen}^\ase(p(e_1, \ldots, e_n)) & = & \emptyset &
\textrm{if $(\den{e_1}^\ase, \ldots, \den{e_n}^\ase) \subseteq \sei(p)$}\\
\mathit{gen}^\ase(e_1 = e_2) & = & \emptyset &
\textrm{if $\den{e_1}^\ase = \den{e_2}^\ase$}\\
\mathit{gen}(\psi_1 \vee \psi_2) & = & 
          \mathit{gen}(\psi_1) &\\
\mathit{gen}(\psi_1 \vee \psi_2) & = & 
          \mathit{gen}(\psi_2) &\\
\mathit{gen}(\psi_1 \wedge \psi_2) & = &
          \mathit{gen}(\psi_1) \cup \mathit{gen}(\psi_2) &\\
\mathit{gen}(\psi_1 \rightarrow \psi_2) & = &
          \mathit{gen}(\psi_2) &\\
\mathit{gen}(\next \psi) & = &
          \emptyset + \mathit{gen}(\psi) &\\
\mathit{gen}(\now{x}{o} . \psi) & = & \{e\} +\ \mathit{gen}(\psi) &
\textrm{if $x \not\in \mathit{fv}(\psi)$, pick an $e \in \term$}\\
&&&\textrm{s.t.\ $\mathit{gen}(\psi[x \mapsto e]) \neq \mathit{err}$}\\
\mathit{gen}(\psi) & = & \mathit{err} & \mathrm{otherwise}\\
\end{array}
$

\noindent
where $\emptyset$ stands for an empty term and indicates that the batch
can be empty.

\end{definition}

Note that this definition interprets conjunctions as unions.
Hence, the formula $\psi \equiv (\now{x}{o} . x = a)\ \wedge (\now{x}{o} . x = b)$
is interpreted as
$\psi \equiv (\now{x}{o} . x \supset \{a\})\ \wedge (\now{x}{o} . x \supset \{b\})$ and
generates a single batch containing $a$ and $b$.

\begin{example}
We present here the generation process for a formula from Example~\ref{ex:formulas}.

\begin{itemize}
\item
$\mathit{gen}(\always{2}(\now{x}{o}. x = b) \rightarrow (\teventually{2} \now{y}{p}. y = a))
=\mathit{gen}(
(\now{x}{o}. x = b) \rightarrow (\now{y}{p}. y = a \vee 
      (\next\nt(\teventually{1} \now{y}{p}. y = a))) \wedge
\next \nt(\always{1}(\now{x}{o}. x = b) \rightarrow (\teventually{2} \now{y}{p}. y = a)))
$ (from Example~\ref{ex:lazy_nt}).

\item
$
\mathit{gen}(
(\now{x}{o}. x = b) \rightarrow (\now{y}{p}. y = a \vee 
      (\next\nt(\teventually{1} \now{y}{p}. y = a))) \wedge
\next \nt(\always{1}(\now{x}{o}. x = b) \rightarrow (\teventually{2} \now{y}{p}. y = a)))
=\\
\mathit{gen}(
(\now{x}{o}. x = b) \rightarrow (\now{y}{p}. y = a \vee 
      (\next\nt(\teventually{1} \now{y}{p}. y = a)))) \cup\\
\mathit{gen}(\next \nt(\always{1}(\now{x}{o}. x = b) \rightarrow (\teventually{2} \now{y}{p}. y = a))) =\\
\printarray[4ex]{{a}} \cup \printarray[4ex]{{\emptyset},{a}} = \printarray[4ex]{{a},{a}}
$
\end{itemize}

Since we have, for the first term of the union:

\begin{itemize}
\item
$
\mathit{gen}(
(\now{x}{o}. x = b) \rightarrow (\now{y}{p}. y = a \vee 
      (\next\nt(\teventually{1} \now{y}{p}. y = a)))) =\\
\mathit{gen}(
\now{y}{p}. y = a \vee 
      (\next\nt(\teventually{1} \now{y}{p}. y = a))) =\\
\mathit{gen}(\now{y}{p}. y = a) =\\
\printarray[4ex]{{a}}
$
\end{itemize}

Similarly we would generate the second term of the union. Note that in both cases
we decided to generate values for the first term of the disjunction. A similar process
can be followed to obtain different values.
\end{example}

\section{sscheck: using $\sstl$ for property-based testing}\label{sec:tool}

We have developed a prototype that allows for using the $\sstl$ logic for property-based testing of Spark Streaming programs, as the Scala library sscheck \cite{sscheckSW}.
This library extends the PBT library ScalaCheck \cite{nilsson2014scalacheck} with custom generators for Spark DStreams and with a property factory that allows 
developers to check a $\sstl$ formula against the finite DStream prefixes generated by another $\sstl$ formula.

\subsection{Design overview\label{subsec:design}}
In order to write a temporal property in sscheck, the user extends the trait (the Scala version of an abstract class) \texttt{DStreamTLProperty}, and then implements some abstract methods to configure Spark Streaming (e.g. defining the batch interval or the Spark master). The method \texttt{DStreamTLProperty.forAllDStream} is used to define temporal ScalaCheck properties:

{\codesize
\begin{verbatim}
  type SSeq[A] = Seq[Seq[A]]
  type SSGen[A] = Gen[SSeq[A]]
  
  def forAllDStream[In:ClassTag,Out:ClassTag](
    generator: SSGen[In])(
    transformation: (DStream[In]) => DStream[Out])(
    formula: Formula[(RDD[In], RDD[Out])])(
    implicit pp1: SSeq[In] => Pretty): Prop 
\end{verbatim}
}
\noindent The function \texttt{forAllDStream} takes a ScalaCheck generator of sequences of sequences of elements, which are interpreted as finite DStream prefixes, so each nested sequence is interpreted as an RDD. Our library defines a case class \texttt{Batch[A]} that extends \texttt{Seq[A]} to represent an RDD for a micro batch, and a case class \texttt{PDStream[A]} that extends \texttt{Seq[Batch[A]]} to represent a finite DStream  prefix. For example \texttt{Batch("scala", "spark")} represents an \texttt{RDD[String]} with 2 elements, and  \texttt{PDStream(Batch("scala", "spark"), Batch(), Batch("spark"))} represents a finite prefix of a \texttt{DStream[String]} consisting of a micro batch with 2 elements, followed by an empty micro batch, and finally a micro batch with a single element. The sscheck classes \texttt{BatchGen} and \texttt{PDStreamGen} and their companion objects can be used to define generators of \texttt{Batch} and \texttt{PDStream} objects using temporal operators, and the trait \texttt{Formula} is used to represent $\sstl$ formulas. See Section \ref{sect:implementation-details} below for details about the user API to write properties with sscheck. 
%
Note the type parameter of \texttt{Formula} is \texttt{(RDD[In], RDD[Out])}, which means in \texttt{formula} the letter corresponding to each instant is a pair of RDDs, one for the input DStream and another for the output DStream. Finally the function \texttt{transformation} is the \emph{test subject} which correctness is checked during the evaluation of the property. 

In order to evaluate the resulting ScalaCheck \texttt{Prop}, first 
we apply a lazy variant of the transformation from Definition~\ref{def:recNextForm} (see Section~\ref{sect:implementation-details} for details.) to \texttt{formula}, in order 
 to get an equivalent formula in next form. Then the following process iterates until the specified number of test cases has passed, or until a failing test case---i.e.\ a counterexample---is found, whatever happens first. A test case of type \texttt{SSeq[In]} is generated using \texttt{generator}, which corresponds to a finite prefix for the input DStream, and a fresh Spark \texttt{StreamingContext} is created. 
The test case, the streaming context, and the transformation are used to create a \texttt{TestCaseContext} that encapsulates the execution of the test case. The program then blocks until the test case is executed completely by the Spark runtime, and then a result for the test case is returned by the test case context. Test case results can be inconclusive, which corresponds to the truth value $?$ in $\sstl$, in case the generated test case is too short for the formula. Internally the test case context defines an input DStream by parallelizing the test case ---using the Spark-testing-base package \cite{KarauTestSpark2015}---, and applies the test subject \texttt{transformation} to it to define an output DStream. It also maintains variables for the number of remaining batches (initialized to the length of the test case), and the current value for the formula, and registers a \texttt{foreachRDD} Spark action that updates the number of remaining batches, and the current formula using the letter simplification procedure from Definition~\ref{def:let_simp}. This action also stops the Spark streaming context once the formula is solved or there are no remaining batches. Other variants of \texttt{forAllDStream} can be used for defining properties with more than one input DStream and one output DStream. 

Therefore \texttt{forAllDStream(gen)(transformation)(formula)} is trying to refute $\forall u_g \in gen(\varphi_g). (\jud{u, 1}{\varphi_p}{\true}{\sse}) \vee (\jud{u, 1}{\varphi_p}{?}{\sse})$ for the Spark interpretation structure $\sse$, formulas $\varphi_g, \varphi_p$ corresponding to \texttt{gen} and \texttt{formula} respectively, and $u \equiv zip(zip(u_g, u_o), u_t)$ where $u_o$ is a word which interpretation under $\sse$ corresponds to the result of applying \texttt{transformation} to the interpretation of $u_g$ under $\sse$, and $u_t = ct\ (ct+b)\ (ct + 2b)\ (ct + 3b)\ldots$ is the sequence of time stamps starting from the unix timestamp $ct$ at the start of the execution of the property and moving $b$ milliseconds at a time for $b$ the configured batch interval. Here $zip$ is the usual operator that combines two sequences element wise to produce a sequence of pairs of elements in the same position, truncating the longest of the two sequences to the length of the shortest. This way we add an additional external universal quantifier on the domain of finite words, as usual in PBT, but inside that scope we have a propositional $\sstl$ formula, and we evaluate the whole formula with the usual sound but incomplete PBT evaluation procedure. 


\subsection{User manual}\label{sect:implementation-details}
In order to check the behaviour of a test subject \texttt{transformation} the user defines a property using $\sstl$ logic by invoking the method \texttt{forAllDStream[In, Out]} with the following arguments:


\begin{description}

\item[Generator \texttt{gen: Gen[Seq[Seq[In]]]}.] It is a regular ScalaCheck generator that produces sequences of sequences of elements, where each nested sequence represents an RDD for a Spark Streaming micro batch, and where the top sequence represent a prefix of a DStream. We represent that with the classes \texttt{Batch[A](points : A*) extends Seq[A]} for the batches, and \texttt{PDStream[A](batches : Batch[A]*) extends Seq[Batch[A]]} for the DStream prefixes. The objects \texttt{PDStreamGen} and \texttt{BatchGen} define functions for a small combinator library for ScalaCheck generators using the temporal operators of $\sstl$. First of all, \texttt{BatchGen.ofN[T](n: Int, g: Gen[T]): Gen[Batch[T]]} can be used to define a batch generator of batches of size $n$ from the elements generated by the ScalaCheck generator $g$. We can then use \texttt{BatchGen.always[A](bg: Gen[Batch[A]], t: Timeout): Gen[PDStream[A]]}, and \texttt{PDStreamGen.always[A](dsg:}\texttt{Gen[PDStream[A]],}\texttt{t: Timeout): Gen[PDStream[A]]} to build more complex generators, using $\sstl$ operators. We also include combinator functions \texttt{next}, \texttt{eventually}, \texttt{until}, and \texttt{release}, that map to $\sstl$ operators in the obvious way. These combinators are also available as methods for the classes \texttt{BatchGen} and \texttt{PDStreamGen}, for defining generators easily with a fluent syntax. 

We currently do not include combinators for the consume operator $\now{x}{o}$, and generators only cover the propositional version of $\sstl$ from \cite{sscheckIFM}. 
Also, just like regular ScalaCheck, we do not include functions for non temporal operators like disjunction, intersection, or implication, and rely on ScalaCheck's \texttt{Gen.oneOf} for implementing the disjunction. 
There are also other combinators to concatenate two \texttt{PDStreamGen} objects, both by concatenating the one \texttt{PDStream} after the other --combinator \texttt{++}--, and by concatenating the \texttt{PDStream} objects batch by batch, in an zip operation--combinator \texttt{+}. Examples~\ref{example-banning}, \ref{ex:countHashtagsOk}, and \ref{ex:sparkTopUntilScalaTop} show the usage of these combinators. 

\item[Test subject \texttt{transformation: (DStream[In]) => DStream[Out]}.] This function that transforms an input DStream into an output DStream is the part of the production code that we are testing with the property. 

\item[Assertion \texttt{formula: (RDD[In], RDD[Out])}.] While the generator defines how to build input DStream prefixes, the assertion formula defines the expected relation between the input DStream and the output DStream. 
It is a value of type \texttt{Formula[(RDD[In], RDD[Out])]}, for \texttt{Formula} a sealed trait that is extended by a case class for each operator in $\sstl$, following the typical implementation of an algebraic data type in Scala. This hierarchy includes the case classes \texttt{Not}, \texttt{Or}, \texttt{And}, \texttt{Implies}, \texttt{Next}, \texttt{Eventually}, \texttt{Always}, \texttt{Until}, and \texttt{Release}, that map to $\sstl$ operators in the obvious way. The consume operator is represented by \texttt{case class BindNext[T](timedAtomsConsumer: TimedAtomsConsumer[T])}, where the class \texttt{TimedAtomsConsumer[T]} just adds a bit on functionality on top of a given \texttt{timedLetterConsumer: Time => T => Formula[T]}, which is a function that defines how to consume the current letter to produce a new formula for the following letter. In this context, \texttt{T} would be equal to \texttt{(RDD[In], RDD[Out])}, containing the value of the input and output micro batches for the current instant, as corresponds to the Spark interpretation structure $\sse$. 
Also, \texttt{case class Solved[T](status : Prop.Status)} represents a solved formula with a value in $\{\true, \false, ?\}$ as correspond to the \texttt{status} value. \texttt{Prop.Status} is a type defined in the ScalaCheck library, that also includes the undefined value, 
and that we use to connect sscheck with ScalaCheck. 
Similarly to what we did for \texttt{PDStream}, the \texttt{Formula} trait and its companion object contain functions and methods \texttt{or}, \texttt{always}, etc, that define a combinator library for formulas. 

Regarding other basic formulas, like predicates and equalities, we can represent them as instances of \texttt{BindNext}, using constant timed atoms consumer functions. Note there is no problem with checking these formulas in the next instant, because timeless formulas have the same truth value at all instants. The combinator library offers a couple of ways to express those applications of \texttt{BindNext} in a nicer way that direct constructor applications, that in particular integrates with the Specs2 matcher assertions that programmers are familiar with. 

\begin{itemize}
\item The first one, used in examples \ref{example-banning}, \ref{ex:getHashtagsOk}, \ref{ex:countHashtagsOk}, \ref{ex:sparkTopUntilScalaTop}, and \ref{ex:alwaysOnlyOneTopHashtag}, is based on the function \texttt{Formula.at[T, A, R <\% Result](proj: (T) => A)(assertion: A => R): Formula[T]}, which builds a formula by composing a projection function on the current letter, for example to extract the input batch, with a function that builds an Spec2 matcher with the result---using Specs2's type \texttt{Result}---, and uses that to build a \texttt{BindNext} instance with a timed atom consumer function that ignores the time argument. In this setting matchers represent predicates and equalities, and regular Scala functions and methods represent $\sstl$ functions, which again corresponds to the Spark interpretation structure $\sse$.  

\item Another option is directly using the function \texttt{Formula.now[T](letterToResult: T => Result): BindNext[T]} that \texttt{Formula.at} uses under the hood, as seen in Example \ref{ex:forallNumRepetitionsLaterCountNumRepetitions}. There is also \texttt{Formula.nowTime[T](letterToResult: (T, Time) => Result): BindNext[T]} for formulas with a time component. 

\item We also have \texttt{Formula.next[T](letterToFormula: T => Formula[T]): Formula[T]} that again builds a \texttt{BindNext} instance with nicer syntax, which is called ``next'' instead of ``now'' because we do not know if the result formula will be timeless, and so the result in the next instant would not necessarily be applicable to the current instant. See e.g. examples \ref{ex:countHashtagsOk}, and \ref{ex:forallNumRepetitionsLaterCountNumRepetitions}. There is also a version \texttt{Formula.nextTime} for using time. In Example~\ref{ex:simpleFormulasPF} we see example usages of \texttt{nowTime} and \texttt{nextTime}. 

\item Finally, functions and methods \texttt{or}, \texttt{always}, etc have variants that accept timed atoms consumer functions, to save some ``now'' and ``next'' applications and write the formula more succinctly, see examples \ref{ex:getHashtagsReferenceImplementationOk}, \ref{ex:hashtagsAreAlwasysCounted}, \ref{ex:alwaysEventuallyZeroCount}, \ref{ex:alwaysPeakImpliesEventuallyTop}, and \ref{ex:forallNumRepetitionsLaterCountNumRepetitions}.

\end{itemize}

\item[Pretty printer witness \texttt{pp1}.] This is just an artifact required by ScalaCheck for printing each generated test case while reporting property evaluation results. ScalaCheck already includes implicit values for most usual types, so passing an explicitly value for this argument is rarely required.

\end{description}

%
Once the generator and the formula are defined, all that is left is using them in a test class that extends Specs2's \texttt{Specification} and sscheck's \texttt{DStreamTLProperty}. A Specs2 test example can be defined using \texttt{DStreamTLProperty.forAllDStream} invoked as \texttt{forAllDStream(gen)(transformation)(formula)}, which returns an object of ScalaCheck's type \texttt{Prop}, that can be used to launch the property check.  



\subsection{Verifying AMP Camp's Twitter tutorial}\label{sect:twitter-example}

In this section we give a flavor of the performance of sscheck on a more complex
example, adapted for Berkeley's AMP Camp training on
Spark,\footnote{\url{http://ampcamp.berkeley.edu/3/exercises/realtime-processing-with-spark-streaming.html}} adding sscheck properties for the functions implemented in that tutorial. The code for these examples is reproduced in \ref{app:ampcamp} and it is available for download 
at \cite{sscheckSWExamples}, while detailed
explanations are available in~\cite{tplpTR}.


Our test subject will work on a stream of \emph{tweets}. A tweet is a piece of text of up 
to 140 characters, together with some meta-information like an identifier for the author 
or the creation date. Those words in a tweet that start with the \verb"#" character are 
called ``hashtags'' and are used by the tweet author to label the tweet, so other users
that later search for tweets with a particular hashtag can locate those related tweets
easily. If many tweets use the same hashtag it becomes  ``popular'' (a so called 
\emph{trending topic}) and can become a topic of discussion between users. For this
reason, Tweeter provides the most popular trending topics in real time, so it is worth
noting that popularity is not measured in absolute terms but in a temporal window 
(that is, it is more popular
a hashtag that appeared 10 times the last minute than one that appeared 20 times
yesterday).
In these examples we check the AMP Camp versions of the functions required to compute
trending topics.
The generators required for checking these functions
consist of a stream of random tweets
containing hashtags from a certain set given as parameter; we will specify, for each
property, the details of the generator. We check the following properties, each corresponding to an Specs2 \cite{etorreborre2015Specs2} example test function. 

\begin{description}
\item[Hashtags correctly extracted (\texttt{getHashtagsOk}).]
We first check whether the hashtags are correctly computed. 
We use a simple generator that \emph{always} generates tweets with hashtags in a predefined set. 
Since we know beforehand
the hashtags in the tweets we check that \emph{always} all the tweets have at least
one hashtag, and the computed hashtags are in the set of hashtags indicated by the argument
given to the generator.

\item[Hashtags correctly counted (\texttt{countHashtagsOk}).]
We also need to make sure that, for a given period (which does not refer to real but
logical time, measured by the number of batches) our functions count all appearances
of hashtags.
In this case our generator first generates only one hashtag (\verb"#spark") and after 
some time only another hashtag (\verb"#scala"), both of them generated with \emph{always}.
For this stream we check that (i) we reach the expected amount of \verb"#spark",
(ii) the amount decreases in the given window \emph{until} it reaches 0, and 
(iii) we reach the expected amount of \verb"#scala".

\item[Trending topic correctly found (\texttt{sparkTopUntilScalaTop}).]
We check now that our functions select as trending topic the hashtag that has
appeared most often and that this trending topic is updated if another one
becomes more popular. We use an \emph{until} generator that produces first tweets
with the hashtags \verb"#scalacheck" and \verb"#spark", using more than twice the latter,
and then it keeps generating tweets with \verb"#scalacheck" and \verb"#scala", but in
this case \verb"#scala" appears more than three times for each \verb"#scalacheck".
The corresponding property checks, also with \emph{until}, that the trending topic
is correctly computed.

\item [There is always exactly one top hashtag (\texttt{alwaysOnlyOneTopHashtag}).] 
Next, we check that there is always just one top hashtag by generating a stream of random tweets and then checking that the output stream of top hashtags always has a single element at each instant in time. This is a simple case of a safety property of the form $\square \mi{result}$.

\item[The count of all hashtags is eventually zero (\texttt{alwaysEventuallyZeroCount}).]
We check that the count for all hashtags reaches eventually zero by using a generator
that creates a stream of tweets that starts with different hashtags and finishes with
tweets without hashtags; this process is repeated inside an \emph{always} operator.
Then, we check that it is always the case that eventually the count for all hashtags 
reaches zero.
Note that this property has the form of a liveness property $\square(\lozenge \mi{result})$.

\item[Periodic trending topic (\texttt{alwaysPeakImpliesEventuallyTop}).]
We generate for a long period random hashtags and a particular hashtag, making sure
the latter happens often. Then we check that, if we reach a peak (e.g.\ 20 appearances
in the window) then it corresponds to the particular hashtag we are generating.
Note that this property has the form of a liveness property $\square(\mi{premise}
\rightarrow \lozenge \mi{result})$.
\end{description}

On Table \ref{fig:tool-props} we present the execution times for these properties, as well as the number of successful generated tests cases used when checking each property. 
The test suite was executed on an Intel Core i7-3517U dual core 1.9 GHz and 8 GB RAM, with Spark running in local mode.  
That is a reasonable time for an integration test, and could be used as an automated validation step in a continuous integration pipeline \cite{fowler2006continuous}. sscheck local execution could be also used for local developing to fix a broken test, using a longer batch interval configuration and smaller number of passing test cases to adapt to an scenario with less computing resources. On the other hand, if a cluster is available, sscheck could be executed using Spark distributed mode ---by setting the \texttt{sparkMaster} field appropriately---, using a shorter batch interval, higher default parallelism, and a higher number of passing tests. In the future we also plan to develop a new feature to allow several test cases for the same property to be execute in parallel. This is not trivial because Spark is limited to a single Spark context per Java virtual machine (JVM) ---see \url{https://issues.apache.org/jira/browse/SPARK-2243}.


\begin{figure}
\begin{tabular}{ l c c }
\hline
Property & Exec.\ time (seconds) & Test cases \\
\hline 
  \texttt{getHashtagsOk} & 46 & 10 \\
  \texttt{countHashtagsOk} & 142 & 15 \\
  \texttt{sparkTopUntilScalaTop} & 56 & 15 \\
  \texttt{alwaysOnlyOneTopHashtag} & 45 & 10 \\
  \texttt{alwaysEventuallyZeroCount} & 187 & 15 \\
  \texttt{alwaysPeakImpliesEventuallyTop} & 303 & 15 \\
\end{tabular}
\caption{AMP Camp's sscheck properties} \label{fig:tool-props}
\end{figure}

\section{Related work}\label{sec:rel}

At first sight, the system presented in this paper can be considered 
an evolution of the data-flow approaches for the verification of reactive systems developed in the past decades, exemplified by systems like Lustre~\cite{halbwachs1992synchronous} and Lutin~\cite{lutin}. 
%
In fact, the idea underlying both stream processing
systems and data-flow reactive systems is very similar: processing a potentially infinite input stream
while generating an output stream. Moreover, 
they usually work with formulas considering both the current state and
the previous ones, which are similar
to the ``forward'' ones presented here.
There are, however, some differences between these two
approaches, being an important one that sscheck is executed in a parallel way using Spark.
%

Lustre is a programming language for reactive systems that is able to verify safety properties by
generating random input streams. The random generation provided by sscheck is 
more refined, since it is possible to define some patterns in the stream in order to verify
some behaviors that can be omitted by purely random generators. Moreover, Lustre specializes
in the verification of critical
systems and hence it has features for dealing with this kind of systems, but lacks
other general features as complex data-structures, although new extensions are included in every new
release. On the other hand, it is not possible to formally verify systems in sscheck;
we focus in a lighter approach for day-to-day programs and, since it
supports all Scala features, its expressive power is greater. 
%
%
%
%
Lutin is a specification language for reactive systems that combines constraints with 
temporal operators. Moreover, it is also possible to generate test cases that depend on the
previous values that the system has generated.
First, these constraints provide more expressive power than the atomic formulas
presented here, and thus the properties stated in Lutin are more expressive than the ones in sscheck.
Although supporting  more expressive formulas would be an interesting subject of future work, in this work we have focused on 
providing a framework where the properties are ``natural'' even for engineers who are
not trained in formal methods; once we have examined the success of this approach we will try to
move into more complex properties.
Second, our framework completely separates the input from the output, and hence it is not possible to
share information between these streams. Although sharing this information is indeed very important
for control systems, we consider that stream processing systems usually deal with external data and
hence this relation is not so relevant for the present tool.
Finally, note that an advantage of sscheck consists in using the same language for both
programming and defining the properties. 

In a similar note, we can consider runtime monitoring of synchronous systems like
\lola~\cite{lola05}, a specification language that allows the user to define properties
in both past and future LTL. \lola\ guarantees bounded memory for monitoring and
allows the user to collect statistics at runtime. On the other hand, as indicated above,
sscheck allows to implement both the programs and the test in the same language and provides
PBT, which simplifies the testing phase, although actual programs cannot be traced.
%


TraceContract \cite{TraceContract} is a Scala library that implements a logic for analyzing
sequences of events (traces). That logic is a hybrid between state machines and temporal logic, 
that is able to express both past time and future time temporal logic formulas, and that supports a form of first 
order quantification over events. The logic is implemented as a shallow internal DSL, just as we do for $\sstl$ in sscheck, 
and it also supports stepwise evaluation of traces so it can be used for online monitoring of a running system, besides
evaluating recorded execution traces. 
On the other hand, TraceContract is not able to generate test cases, and it is not integrated with
any standard testing library like Specs2.

Regarding testing tools for Spark, the most clear precedent is the unit test framework
Spark Test Base~\cite{KarauTestSpark2015b}, which also integrates ScalaCheck for
Spark but only for Spark core. To the best of out knowledge, there is no previous library supporting
property-based testing for Spark Streaming. 

\section{Conclusions and future work}\label{sec:conc}

In this paper we have presented sscheck, a property-based testing tool for Spark Streaming programs.
sscheck allows the user to define generators and to state properties using $\sstl$, an extension
of Linear Temporal Logic with timeouts in temporal operators and a special operator for binding the
current batch and time. This logic allows us to define a stepwise transformation that only requires/generates
the current batch; using this feature the Scala implementation of sscheck takes advantage of
lazy functions to efficiently implement the tool. The benchmarks presented in the paper show that
the approach works well in practice.
With these features in mind, we hope sscheck will be accepted by the industry; we consider the
presentation at Apache Europe~\cite{riesco2016apacheTemporal} and citations in books written by remarkable members of the Spark community \cite{karau2017high} are important steps in this direction.

There are many open lines of future work.
First, adding support for arbitrary nesting of ScalaCheck \texttt{forall} and \texttt{exists} quantifiers inside $\sstl$ formula 
would be an interesting extension. Moreover, we also consider developing versions for other languages with Spark API, in particular Python, or supporting other SPS, like Apache Flink \cite{carbone2015lightweight} or Apache Bean \cite{akidau2015dataflow}. This would require novel solutions, as these systems are not based on synchronous micro-batching but they process events one at a time, and also have interesting additional features like the capability for handling different event time characteristics for supporting out of order streams, and several types of windows \cite{akidau2015dataflow}. 
%
Besides, we plan to explore whether the execution of several test cases in parallel minimize the test suite execution time. 
We could also improve the sscheck library interface, employing advanced Scala DSL techniques like the Magnet Pattern \cite{sprayMagnet2012} to make formulas easier to write and read. 
%
%
Finally, we intend to explore other formalisms for expressing temporal and cyclic behaviors~\cite{DBLP:journals/iandc/Wolper83}.

{

}

\appendix

\section{Proofs\label{app:proofs}}

We present in this sections the proofs for the theorems presented in the paper.

\noindent
\textbf{Theorem \ref{th:t_eq}.}
\textit{Given a formula $\varphi \in \sstl$ such that $\varphi$ does not contain variables in temporal
connectives, we have $\nt(\varphi) = \nt^e(\varphi)$.}
\begin{proof}
We prove it by induction on the structure of the formula. The base cases are the formulas
for $\true$, $\false$, terms, atomic propositions, and equalities, that are not modified and hence
the property holds.

Then, it is easy to see that the property holds for the formulas defining and, or, implication, next,
and consume just by applying the induction hypothesis, since both functions
apply the same transformation.

Finally, we need to apply induction on the time used in temporal connectives. We present the
proof for the always connective; the rest of them follow the same schema. For the base case
we have:

\begin{itemize}
\item
$\nt(\always{1} \varphi) = \nt(\varphi)$.

\item
$\nt^e(\always{1} \varphi) = 
          \nt^e(\varphi) \wedge \next^{0} \nt^e(\varphi) = \nt^e(\varphi)$
\end{itemize}

This case holds by induction hypothesis in the structure of the formula. Then, assuming
$\nt(\teventually{t} \varphi) = \nt^e(\teventually{t} \varphi)$, with $t \geq 2$, we need to
prove $\nt(\teventually{t+1} \varphi) = \nt^e(\teventually{t+1} \varphi)$.

$$
\begin{array}{lll}
\nt(\always{t + 1} \varphi) & = &
          \nt(\varphi) \wedge \next \nt(\always{t} \varphi)\\
          & =^{\mi{HI}} &  \nt(\varphi) \wedge \next \nt^e(\always{t} \varphi)\\
          & =^{\mi{HI (struct)}} &  \nt^e(\varphi) \wedge \next \nt^e(\always{t} \varphi)\\
          & =^{\mi{def}} &  \nt^e(\varphi) \wedge \next (\nt^e(\varphi) \wedge \next \nt^e(\varphi) \wedge \ldots \wedge 
          \next^{t-1} \nt^e(\varphi))\\
          & = &
          \nt^e(\varphi) \wedge \next \nt^e(\varphi) \wedge \ldots \wedge 
          \next^{t} \nt^e(\varphi)\\
\end{array}
$$
\end{proof}

\noindent
\textbf{Lemma~\ref{lemma:n}.}
\textit{
Given an alphabet $\Sigma$ and
formulas $\varphi, \varphi' \in \sstl$,
if $\forall u \in \Sigma^*. u,1 \vDash \varphi \iff u,1 \vDash \varphi'$
then $\forall u \in \Sigma^*, \forall n \in \mathbb{N}^+. u,n \vDash \varphi \iff u,n \vDash \varphi'$.
}
\begin{proof}
Since $u\equiv a_1\ldots a_m$, $m \in \mathbb{N}$, we distinguish the cases $n > m$ and
$n \leq m$:
\begin{description}
\item[$n > m$] It is easy to see for all possible formulas that only ? can be obtained,
so the property trivially holds.

\item[$n \leq m$] Then we have $u' \equiv a_n\ldots a_m$ and, since we know that 
$u' \vDash \varphi \iff u' \vDash \varphi'$, the property holds.
\end{description}
\end{proof}

\noindent
\textbf{Theorem \ref{th:equiv_next}.}
\textit{Given an alphabet $\Sigma$,
an interpretation $\ase$, and
formulas $\varphi, \varphi' \in \sstl$, such that $\varphi' \equiv \nt(\varphi)$,
we have 
$\forall u \in (\Sigma \times \mathbb{N})^*. u \vDash^{\ase} \varphi \iff u \vDash^{\ase} \varphi'$.}

\begin{proof}
We apply induction on formulas.

\textbf{Base case.} It is straightforward to see that the result holds for the constants $\true$
and $\false$ and for an atomic predicate $p$.

\textbf{Induction hypothesis.} Given the formulas $\varphi_1, \varphi_2, \varphi'_1, 
\varphi'_2 \in sstl$, such that $\varphi'_1 \equiv \nt(\varphi_1)$ and
$\varphi'_2 \equiv \nt(\varphi_2)$, we have
$\forall u \in \Sigma^*. u \vDash \varphi_i \iff u \vDash \varphi'_i$, $i \in \{1,2\}$.

\textbf{Inductive case.} We distinguish the different formulas in $\sstl$:
\begin{itemize}
\item
For the formulas $\false, \true, p, \neg \varphi_1, \varphi_1 \vee \varphi_2,
\varphi_1 \wedge \varphi_2$, and $\varphi_1 \rightarrow \varphi_2$ is straightforward to see that the result holds, since the same operators are kept and the subformulas are equivalent by
hypotheses.

\item
For the formula $t_1 = t_2$ is straightforward to see that the result holds, since it remains
unchanged.

\item
For the formula $\now{x}{o} . \varphi$ is also straightforward, since by hypothesis the
subformula is equivalent and then the same variables are bound.

\item
Given the formula $\next \varphi_1$, we have to prove that 
$\forall u \in \Sigma^*. u \vDash \next \varphi_1 \iff u \vDash \next \varphi'_1$. This
expression can be transformed using the definition for the satisfaction for the next
operator into $\forall u \in \Sigma^*. u, 2 \vDash \varphi_1 \iff u, 2 \vDash \varphi'_1$,
which holds by hypothesis and Lemma~\ref{lemma:n}.

\item
Given the formula $\teventually{t} \varphi_1$, $t \in \mathbb{N}^+$, we have to prove that
$\forall u \in \Sigma^*. u \vDash \teventually{t} \varphi_1 \iff 
u \vDash  \varphi_1' \vee \next \varphi_1' \vee \ldots \vee 
          \next^{t-1} \varphi_1'$.
We distinguish the possible values for $u \vDash \teventually{t} \varphi_1$:
\begin{itemize}
\item
$u \vDash \teventually{t} \varphi_1 : \top$. In this case the property holds because
there exists $i$, $1 \leq i \leq t$ such that $u, i \vDash \varphi_1 : \top$. Hence,
$u \vDash \next^{i-1} \varphi'_1$ by hypothesis and the definition of the
next operator (note that for $i = 1$ we just have $u \vDash \varphi'$).

\item
$u \vDash \teventually{t} \varphi_1 : \bot$. In this case $\forall i, 1 \leq i \leq t,
u, i \vDash \varphi_1 : \bot$, so we have $u \vDash \next^{i-1} \varphi'_1 : \bot$ for
$1 \leq i \leq t$ and the transformation is also evaluated to $\bot$.

\item
$u \vDash \teventually{t} \varphi_1 :\ ?$. In this case we have $u$ of length $n$,
$n < t$, and $\forall i, 1 \leq i \leq n, u, i \vDash \varphi_1 : \bot$. Hence, we
have $u \vDash \next^{i-1} \varphi'_1 : \bot$ for $1 \leq i \leq n$ and 
$u \vDash \next^{j-1} \varphi'_1 :\ ?$ for $n + 1 \leq j \leq t$. Hence, we have
$\bot\ \vee\ \ldots\ \vee\ \bot\ \vee\ ?\ \vee\ \ldots\ \vee\ ? =\ ?$ and the property holds.
\end{itemize}

\item
The analysis for $\always{t} \varphi_1$ is analogous to the one for 
$\teventually{t} \varphi_1$.

\item
Given the formula $\varphi_1 \tuntil{t} \varphi_2$, $t \in \mathbb{N}^+$,
we have to prove that
$\forall u \in \Sigma^*. u \vDash \varphi_1 \tuntil{t} \varphi_2 \iff 
u \vDash
\varphi'_2 \vee (\varphi'_1 \wedge \next \varphi'_2)
          \vee \ldots \vee
          (\varphi'_1 \wedge \next \varphi'_1 \wedge \ldots \wedge
          \next^{t-2} \varphi'_1 \wedge \next^{t-1} \varphi'_2)$.
We distinguish the possible values for $u \vDash \varphi_1 \tuntil{t} \varphi_2$:
\begin{itemize}
\item
$u \vDash \varphi_1 \tuntil{t} \varphi_2 : \top$. In this case we have from the definition
that $\exists i, 1 \leq i \leq t$ such that $u, i \vDash \varphi_2 : \top$ and
$\forall j, 1 \leq j < i, u, j \vDash \varphi_1 : \top$. Hence, applying the induction
hypothesis we have
$u \vDash \varphi'_1 \wedge \next \varphi'_1 \wedge \ldots \next^{i-2} \varphi'_1 \wedge
\next^{i-1} \varphi'_2 : \top$, and hence the property holds.

\item
$u \vDash \varphi_1 \tuntil{t} \varphi_2 : \bot$.
\begin{itemize}
\item
Case a) $\forall i, 1 \leq i \leq t. u, i \vDash \varphi_2 : \bot$. In this case we have
$\forall i, 1 \leq i \leq t, u, i \vDash \next^{i-1} \varphi'_2 : \bot$, and hence the
complete formula is evaluated to $\bot$.

\item
Case b) $\exists i, 1 \leq i \leq t, \forall j, 1 < j \leq i. u, j \vDash \varphi_1 : \top$,
$u, j \vDash \varphi_2 : \bot$
$u, i \vDash \varphi_1 : \bot$, and $u, i \vDash \varphi_2 : \bot$. In this case we have
$\forall k, 0 \leq k < i, u \vDash \next^{k} \varphi'_2 : \bot$ and
$u \vDash \next^{i-1} \varphi'_1 : \bot$ by inductive hypothesis. Hence, all the conjunctions
are evaluated to $\bot$ and the property holds.
\end{itemize}

\item
$u \vDash \varphi_1 \tuntil{t} \varphi_2 :\ ?$. In this case we have $u$ of length $n$,
$n < t$, $\forall i, 1 \leq i \leq n, u, i \vDash \varphi_2 : \bot$, and
$u, i \vDash \varphi_1 : \top$. Hence, the first $i$ conjunctions in the transformation
are evaluated to $\bot$ by the induction hypothesis, while the rest are evaluated to ?
by the definition of the next operator and the property holds.
\end{itemize}

\item
The analysis for $\varphi_1 \trelease{t} \varphi_2$ is analogous to the one for 
$\varphi_1 \tuntil{t} \varphi_2$, taking into account that formula also holds if
$\varphi_2$ always holds.
\end{itemize}
\end{proof}


\section{Introduction to Spark and Spark Streaming \label{sect:spark-intro}}
Spark \cite{zaharia2012resilient} is a distributed processing engine that was designed as 
an alternative to Hadoop MapReduce~\cite{marz2015big}, but with a focus on iterative 
processing---e.g.\ to implement distributed machine learning algorithms---and interactive 
low latency jobs---e.g.\ for ad hoc SQL queries on massive datasets. The key to achieving 
these goals is an extended memory hierarchy that allows for an increased performance in many situations, and a data model based on immutable collections inspired in functional programming that is the basis for its fault tolerance mechanism. The core of Spark is a batch computing framework \cite{zaharia2012resilient} that is based on manipulating so called Resilient Distributed Datasets (RDDs), which provide a fault tolerant implementation of distributed collections. Computations are defined as transformations on RDDs, that should be deterministic and side-effect free, as the fault tolerance mechanism of Spark is based on its ability to recompute any fragment (partition) of an RDD when needed. Hence Spark programmers are encouraged to define RDD transformations that are pure functions from RDD to RDD, and the set of predefined RDD transformations includes typical higher-order functions like map, filter, 
etc., as well as aggregations by key and joins for RDDs of key-value pairs.  
We can also use Spark actions, which allow us to collect results into the \emph{driver program} or store them into an external data store. The driver program is the local process that starts the connection to the Spark cluster, and issues the execution of Spark jobs, acting as a client of the Spark cluster. Spark actions are impure, so idempotent actions are recommended in order to ensure a deterministic behavior even in the presence of recomputations triggered by the fault tolerance or speculative task execution 
mechanisms \cite{sparkProgrammingGuide}. 
Spark is written in Scala and offers APIs for Scala, Java, Python, and R; in this work we focus on the Scala API. The example in Figure~\ref{fig:letters1} uses the Scala Spark shell to implement a variant of the famous word count example that in this case computes the number of occurrences of each character in a sentence. For that we use \texttt{parallelize}, a feature of Spark that allows us to create an RDD from a local collection, which is useful for testing. We start with a set of chars distributed among $3$ partitions,
we pair each char with a 1 by using \texttt{map}, and then group by first component in the pair and sum by the second by using
\texttt{reduceByKey} and the addition function (\verb"_+_"), thus obtaining a set of (char, frequency) pairs. We collect this set into an \texttt{Array} in the driver with \texttt{collect}.

\begin{figure}[t]
{\footnotesize
\begin{verbatim}
scala> val cs: RDD[Char] = sc.parallelize("let's count some letters", numSlices=3)
scala> cs.map{(_, 1)}.reduceByKey{_+_}.collect()¡
res4: Array[(Char, Int)] = Array((t,4), ( ,3), (l,2), (e,4), (u,1), (m,1), (n,1), 
                                 (r,1), (',1), (s,3), (o,2), (c,1))
\end{verbatim}
}
\caption{Letter count in Spark\label{fig:letters1}}
\end{figure}

Besides the core RDD API, the Spark release contains a set of high level libraries that accelerates the development of Big Data processing applications, and that are also one of the reasons for its growing popularity. This includes libraries for scalable machine learning, graph processing, a SQL engine, and Spark Streaming, which is the focus of this work. 
In Spark Streaming, the notions of transformations and actions are extended from RDDs to DStreams (Discretized Streams), which are series of RDDs corresponding to splitting an input data stream into fixed time windows, also called micro batches. Micro batches are generated at a fixed rate according to the configured \emph{batch interval}. Spark Streaming is synchronous in the sense that given a collection of input and transformed DStreams, all the batches for each DStream are generated at the same time as the batch interval is met. Actions on DStreams are also periodic and are executed synchronously for each micro batch. The code in Figure~\ref{fig:letters2} is
the streaming version of the code in Figure~\ref{fig:letters1}. In this case we process a DStream of characters, where batches are obtained by splitting a String into pieces by
making groups (RDDs) of $4$ consecutive characters. 
We use the testing 
utility class \texttt{QueueInputDStream}, which generates batches by picking RDDs from a queue,
to generate the input DStream by parallelizing each substring into an RDD with $3$ partitions.  
The program is executed using the local master mode of Spark,
which replaces slave nodes in a distributed cluster by local threads, which is useful for developing and testing.

\begin{figure}[t]
\begin{tabular}{ll}
\begin{minipage}{0.72\textwidth}
{\scriptsize
\begin{verbatim}
object HelloSparkStreaming extends App {
  val conf = new SparkConf().setAppName("HelloSparkStreaming")
                            .setMaster("local[5]")
  val sc = new SparkContext(conf)
  val batchInterval = Duration(100)
  val ssc = new StreamingContext(sc, batchInterval)
  val batches = "let's count some letters, again and again"
                .grouped(4)
  val queue = new Queue[RDD[Char]]
  queue ++= batches.map(sc.parallelize(_, numSlices = 3))
  val css : DStream[Char] = ssc.queueStream(queue, 
                                            oneAtATime = true)
  css.map{(_, 1)}.reduceByKey{_+_}.print()
  ssc.start()
  ssc.awaitTerminationOrTimeout(5000)
  ssc.stop(stopSparkContext = true)
}
\end{verbatim}
}
\end{minipage} &
\begin{minipage}{0.25\textwidth}
%
{\scriptsize
\begin{verbatim}
-----------------------
Time: 1449638784400 ms
-----------------------
(e,1)
(t,1)
(l,1)
(',1)
...
-----------------------
Time: 1449638785300 ms
-----------------------
(i,1)
(a,2)
(g,1)
-----------------------
Time: 1449638785400 ms
-----------------------
(n,1)
\end{verbatim}
}
\end{minipage}

\end{tabular}
\caption{Letter count in Spark Streaming\label{fig:letters2}}
\end{figure}


\section{Overview of property-based testing and ScalaCheck \label{sect:pbt-intro}}
Classical unit testing with xUnit-like frameworks \cite{meszaros2007xunit} is based on specifying input -- expected output pairs, and then comparing the expected output with the actual output obtained by applying the test subject to the input. 
On the other hand, in property-based testing (PBT) a test is expressed as a property, which is a formula in a restricted version of first order logic that relates program input and output. The testing framework checks the property by evaluating it against a bunch of randomly generated inputs. If a counterexample 
for the property is found then the test fails, otherwise it passes. This allows
developers to obtain quite a good test coverage of the production code with a fairly small investment on development time, specially when compared to xUnit frameworks. However xUnit frameworks are still useful for testing corner cases that would be difficult to cover with a PBT property. 
The following is a ``hello world'' ScalaCheck property that checks the commutativity of addition:\footnote{Here we use the integration of ScalaCheck with the Specs2 \cite{etorreborre2015Specs2} testing library.}

{\small
\begin{verbatim}
class HelloPBT extends Specification with ScalaCheck {
  def is = s2"""Hello world PBT spec, 
                where int addition is commutative $intAdditionCommutative"""
                
  def intAdditionCommutative = 
    Prop.forAll("x" |: arbitrary[Int], "y" |: arbitrary[Int]) { (x, y) => 
      x + y === y + x
  }.set(minTestsOk = 100) 
}
\end{verbatim}
}

PBT is based on \emph{generators} (the functions in charge of computing the inputs,
which define the domain of discourse for a formula) and
\emph{assertions} (the atoms of a formula), which together with a \emph{quantifier} form a \emph{property} (the formula to be checked). In the example above the universal quantifier \texttt{Prop.forAll} is used to define a property that checks whether the assertion \texttt{x + y === y + x} holds for 100 values for \texttt{x} and \texttt{y} randomly generated by two instances of the integer generator \texttt{arbitrary[Int]}. 
Each of those pairs of values generated for \texttt{x} and \texttt{y} is called a \emph{test case}, and a test case that refutes the assertions of a property is called a \emph{counterexample}. Here \texttt{arbitrary} is a higher order generator that is able to generate random values for 
predefined and custom types. 
Besides universal quantifiers, ScalaCheck supports existential quantifiers---although these are not much used in practice~\cite{nilsson2014scalacheck,venners2015Exists}---, and logical operators to compose properties.  
PBT is a sound procedure to check the validity of the formulas implied by the properties,  in the sense that any counterexample that is found can be used to build a definitive proof that the property is false. However, it is not complete, as there is no guarantee that the whole space of test cases is explored exhaustively, so if no counterexample is found then we cannot conclude that the property holds for all possible test cases that could had been generated: \emph{all failing properties are definitively false, but not all passing properties are definitively true}.
PBT is a lightweight approach that does not attempt to perform sophisticated automatic deductions, but it provides a very fast test execution that is suitable for the test-driven development (TDD) cycle, and empirical studies \cite{claessen2011quickcheck,shamshiri2015random} have shown that in practice random PBT obtains good results, with a quality comparable to more sophisticated techniques. This goes in the line of assuming that in general testing of non trivial systems is often incomplete, as the effort of completely modeling all the possible behaviors of the system under test with test cases is not cost effective in most software development projects, except for critical systems. 

\section{Code for AMP Camp's Twitter tutorial with sscheck}\label{app:ampcamp}

Now we will present a more complex example, adapted for Berkeley's AMP Camp training on
Spark,\footnote{\url{http://ampcamp.berkeley.edu/3/exercises/realtime-processing-with-spark-streaming.html}} but adding sscheck properties for the functions implemented in that tutorial. The complete code for these examples is available at \url{https://github.com/juanrh/sscheck-examples/releases/tag/0.0.4}.
 
Our test subject will be an object \texttt{TweetOps}, which defines a series of operations on a stream of tweets.  A tweet is a piece of text of up to 140 characters, together with some meta-information like an identifier for the author or the creation date. Those words in a tweet that start with the \verb"#" character are called ``hashtags'', and are used by the tweet author to label the tweet, so other users that later search for tweets with a particular hashtag might locate those related tweets easily. The operations below take a stream of tweets and, respectively, generate the stream for the set of hashtags in all the tweets; the stream of pairs (hashtags, number of occurrences) in a sliding time window with the specified size\footnote{See \url{https://spark.apache.org/docs/1.6.2/streaming-programming-guide.html\#window-operations} for details on Spark Streaming window operators.}; and the stream that contains a single element for the most popular hashtag, i.e. the hashtag with the highest number of occurrences, again for the specified time window. 

{\codesize
\begin{verbatim}
object TweetOps {
 def getHashtags(tweets: DStream[Status]): DStream[String]
 def countHashtags(batchInterval: Duration, windowSize: Int)
                  (tweets: DStream[Status]): DStream[(String, Int)]
 def getTopHashtag(batchInterval: Duration, windowSize: Int)
                  (tweets: DStream[Status]): DStream[String]
}
\end{verbatim}
}

\noindent 
In this code, the class \texttt{twitter4j.Status} from the library Twitter4J~\cite{yamamoto2010twitter4j} is used to represent each particular tweet. In the original AMP Camp training, the class \texttt{TwitterUtils}\footnote{\url{https://spark.apache.org/docs/1.6.0/api/java/org/apache/spark/streaming/twitter/TwitterUtils.html}} is used to define a \texttt{DStream[Status]} by repeatedly calling the Twitter public API to ask for new tweets. Instead, in this example we replace the Twitter API by an input DStream defined by using an sscheck generator, so we can control the shape of the tweets that will be used as the test inputs. To do that we employ the mocking \cite{mackinnon2001endo} library Mockito \cite{kaczanowski2012practical} for stubbing \cite{fowler2007mocks} \texttt{Status} objects, i.e. to easily synthetize objects that impersonate a real \texttt{Status} object, and that provide predefined answers to some methods, in this case the method that returns the text for a tweet. 

{\codesize
\begin{verbatim}
object TwitterGen {
  /** Generator of Status mocks with a getText method
   *  that returns texts of up to 140 characters
   *  
   *  @param noHashtags if true then no hashtags are generated in the 
   *  tweet text
   * */
  def tweet(noHashtags: Boolean = true): Gen[Status]
  /** Take a Status mocks generator and return a Status mocks 
   *  generator that adds the specified hashtag to getText
   * */
  def addHashtag(hashtagGen: Gen[String])
                (tweetGen: Gen[Status]): Gen[Status]
  def tweetWithHashtags(possibleHashTags: Seq[String]): Gen[Status] 
  def hashtag(maxLen: Int): Gen[String]
  def tweetWithHashtagsOfMaxLen(maxHashtagLength: Int): Gen[Status]
}
\end{verbatim}
}

\subsection{Extracting hashtags}
Now we are ready to write our first property, which checks that \texttt{getHashtags} works correctly, that is, it computes the set of \emph{hashtags} (words
starting with \verb"#"). 
In the property we generate tweets that use a predefined set
of hashtags, and then we check that all hashtags produced in the output are contained
in that set.

\begin{example}\label{ex:getHashtagsOk}
{\codesize
\begin{verbatim}
def getHashtagsOk = {
  type U = (RDD[Status], RDD[String])
  val hashtagBatch = (_ : U)._2
  
  val numBatches = 5
  val possibleHashTags = List("#spark", "#scala", "#scalacheck")
  val tweets = BatchGen.ofNtoM(5, 10, 
                              tweetWithHashtags(possibleHashTags)
                              )
  val gen = BatchGen.always(tweets, numBatches)
  
  val formula = always { 
    at(hashtagBatch){ hashtags =>
      hashtags.count > 0 and
      ( hashtags should foreachRecord(possibleHashTags.contains(_)) ) 
    }
  } during numBatches
 
  forAllDStream(
    gen)(
    TweetOps.getHashtags)(
    formula)
}
\end{verbatim}
}
\end{example}

In the next example we use the ``reference implementation'' PBT technique \cite{nilsson2014scalacheck} to check the implementation of \texttt{TweetOps.getHashtags}, which is based on the Spark transformations \texttt{flatMap} and \texttt{filter} also using \texttt{String.startsWith}, against a regexp-based reference implementation. This gives us a more thorough test, because we use a different randomly generated set of hashtags for each batch of each test case, instead of a predefined set of hashtags for all the test cases.

\begin{example}\label{ex:getHashtagsReferenceImplementationOk}
{\codesize
\begin{verbatim}
private val hashtagRe = """#\S+""".r
private def getExpectedHashtagsForStatuses(statuses: RDD[Status])
: RDD[String] = 
  statuses.flatMap { status => hashtagRe.findAllIn(status.getText)}

def getHashtagsReferenceImplementationOk = {
  type U = (RDD[Status], RDD[String])    
  val (numBatches, maxHashtagLength) = (5, 8)

  val tweets = BatchGen.ofNtoM(5, 10, 
                               tweetWithHashtagsOfMaxLen(maxHashtagLength))                            
  val gen = BatchGen.always(tweets, numBatches)
  
  val formula = alwaysR[U] { case (statuses, hashtags) => 
    val expectedHashtags = getExpectedHashtagsForStatuses(statuses).cache()
    hashtags must beEqualAsSetTo(expectedHashtags)
  } during numBatches

  forAllDStream(
    gen)(
    TweetOps.getHashtags)(
    formula)
}
\end{verbatim}
}
\end{example}

\subsection{Counting hashtags}

In order to check \texttt{countHashtags}, in the following property we setup a scenario where the hashtag \verb"#spark" is generated for some period, and then the hashtag \verb"#scala" is generated for another period, 
and we express the expected counting behaviour with several subformulas: we expect to get the expected count of hashtags for spark for the first period (\texttt{laterAlwaysAllSparkCount}); we expect to eventually get the expected count of hastags for scala (\texttt{laterScalaCount}); and we expect that after reaching the expected count for spark hashtags, we would then decrease the count as time passes and elements leave the sliding window (\texttt{laterSparkCountUntilDownToZero}). 

\begin{example}\label{ex:countHashtagsOk}
{\codesize
\begin{verbatim}
def countHashtagsOk = {
  type U = (RDD[Status], RDD[(String, Int)])
  val countBatch = (_ : U)._2
  
  val windowSize = 3
  val (sparkTimeout, scalaTimeout) = (windowSize * 4, windowSize * 2)
  val sparkTweet = tweetWithHashtags(List("#spark"))
  val scalaTweet = tweetWithHashtags(List("#scala"))
  val (sparkBatchSize, scalaBatchSize) = (2, 1)
  val gen = BatchGen.always(BatchGen.ofN(sparkBatchSize, sparkTweet), 
                                         sparkTimeout) ++  
            BatchGen.always(BatchGen.ofN(scalaBatchSize, scalaTweet), 
                                         scalaTimeout)
  
  def countNHashtags(hashtag : String)(n : Int)  = 
    at(countBatch)(_ should existsRecord(_ == (hashtag, n : Int)))
  val countNSparks = countNHashtags("#spark") _
  val countNScalas = countNHashtags("#scala") _
  val laterAlwaysAllSparkCount =  
    later { 
        always { 
          countNSparks(sparkBatchSize * windowSize)
        } during (sparkTimeout -2) 
    } on (windowSize + 1) 
  val laterScalaCount = 
    later { 
      countNScalas(scalaBatchSize * windowSize)
    } on (sparkTimeout + windowSize + 1)
  val laterSparkCountUntilDownToZero = 
    later { 
      { countNSparks(sparkBatchSize * windowSize) } until {
        countNSparks(sparkBatchSize * (windowSize - 1)) and
          next(countNSparks(sparkBatchSize * (windowSize - 2))) and
          next(next(countNSparks(sparkBatchSize * (windowSize - 3)))) 
        } on (sparkTimeout -2) 
    } on (windowSize + 1)
  val formula = 
    laterAlwaysAllSparkCount and 
    laterScalaCount and 
    laterSparkCountUntilDownToZero

  forAllDStream(
    gen)(
    TweetOps.countHashtags(batchInterval, windowSize)(_))(
    formula)
}
\end{verbatim}
}
\end{example}


Then we check the safety of \texttt{countHashtags} by asserting that any arbitrary generated hashtag is never skipped in the count. Here we again exploit the reference implementation technique to extract the expected hashtags, and join this with the output counts, so we can assert that all and only all expected hastags are counted, and that those countings are never zero at the time the hashtag is generated. 

\begin{example}\label{ex:hashtagsAreAlwasysCounted}
{\codesize
\begin{verbatim}
def hashtagsAreAlwasysCounted = {
  type U = (RDD[Status], RDD[(String, Int)])
  val windowSize = 3
  val (numBatches, maxHashtagLength) = (windowSize * 6, 8)
  
  val tweets = BatchGen.ofNtoM(5, 10, 
                               tweetWithHashtagsOfMaxLen(maxHashtagLength))      
  val gen = BatchGen.always(tweets, numBatches)
      
  val alwaysCounted = alwaysR[U] { case (statuses, counts) =>  
    val expectedHashtags = getExpectedHashtagsForStatuses(statuses).cache()
    val expectedHashtagsWithActualCount = 
      expectedHashtags
       .map((_, ()))
       .join(counts)
       .map{case (hashtag, (_, count)) => (hashtag, count)}
       .cache()
    val countedHashtags = expectedHashtagsWithActualCount.map{_._1}
    val countings = expectedHashtagsWithActualCount.map{_._2}
    
    // all hashtags have been counted
    countedHashtags must beEqualAsSetTo(expectedHashtags) and
    // no count is zero
    (countings should foreachRecord { _ > 0 }) 
  } during numBatches
  
  forAllDStream(
    gen)(
    TweetOps.countHashtags(batchInterval, windowSize)(_))(
    alwaysCounted)

} 
\end{verbatim}
}
\end{example}

\subsubsection{Getting the most popular hashtag}



Now we check the correctness of \texttt{getTopHashtag}, that extracts the most ``popular'' hashtag, i.e. the hashtag with the highest number of occurrences at each time window. For that we use the following property where we define a scenario in which we start with the hashtag \verb"#spark" as the most popular (generator \texttt{sparkPopular}), and after that the hashtag \verb"#scala" becomes the most popular (generator \texttt{scalaPopular}), and asserting on the output DStream that \verb"#spark" is the most popular hashtag until \verb"#scala" is the most popular.

\begin{example}\label{ex:sparkTopUntilScalaTop}
{\codesize
\begin{verbatim}
def sparkTopUntilScalaTop = {
  type U = (RDD[Status], RDD[String])
  
  val windowSize = 1
  val topHashtagBatch = (_ : U)._2
  val scalaTimeout = 6
  val sparkPopular = 
    BatchGen.ofN(5, tweetWithHashtags(List("#spark"))) +
    BatchGen.ofN(2, tweetWithHashtags(List("#scalacheck"))) 
  val scalaPopular = 
    BatchGen.ofN(7, tweetWithHashtags(List("#scala"))) +
    BatchGen.ofN(2, tweetWithHashtags(List("#scalacheck"))) 
  val gen = BatchGen.until(sparkPopular, scalaPopular, scalaTimeout) 
    
  val formula = 
    { at(topHashtagBatch)(_ should foreachRecord(_ == "#spark" )) } until {
      at(topHashtagBatch)(_ should foreachRecord(_ == "#scala" ))
    } on (scalaTimeout)
  
  forAllDStream(
    gen)(
    TweetOps.getTopHashtag(batchInterval, windowSize)(_))(
    formula)
}
\end{verbatim}
}
\end{example}

Finally, we state the safety of \texttt{getTopHastag} by checking that there is always one top hashtag.

\begin{example}\label{ex:alwaysOnlyOneTopHashtag}
{\codesize
\begin{verbatim}
def alwaysOnlyOneTopHashtag = {
  type U = (RDD[Status], RDD[String])
  val topHashtagBatch = (_ : U)._2

  val (numBatches, maxHashtagLength) = (5, 8)   
  val tweets = 
    BatchGen.ofNtoM(5, 10, 
                    tweetWithHashtagsOfMaxLen(maxHashtagLength))

  val gen = BatchGen.always(tweets, numBatches)    
  val formula = always { 
    at(topHashtagBatch){ hashtags =>
      hashtags.count === 1 
    }
  } during numBatches
  
  forAllDStream(gen)(
    TweetOps.getTopHashtag(batchInterval, 2)(_))(
    formula)
}
\end{verbatim}
}
\end{example}

\subsubsection{Defining liveness properties with the consume operator}\label{sect:consume-liveness}

So far we have basically defined two types of properties: properties where we simulate a particular scenario, and safety properties where we assert that we will never reach a particular ``bad'' state. It would be also nice to be able to write liveness properties in sscheck, which is another class of properties typically used with temporal logic, where we express that something good keeps happening with a formula of the shape of $\always{t_1} (\varphi_1 \rightarrow \teventually{t_2} \varphi_2)$. In this kind of formulas it would be useful to define the conclusion formula $\varphi_2$ that should happen later, based on the value of the word that happened when the premise formula $\varphi_1$ was evaluated. This was our motivation for adding to the $\sstl$ logic the consume operator $\now{x}{o}.\varphi$, that can be used in liveness formulas of the shape $\always{t_1} (\now{x}{o}.\teventually{t_2} \varphi_2)$ or $\always{t_1} (\now{x}{o}.\varphi_1 \rightarrow \teventually{t_2} \varphi_2)$. One example of the former is the following liveness property for \texttt{countHashtags}, that checks that always each hashtag eventually gets a count of 0, if we generate empty batches at the end of the test case so all hashtags end up getting out of the counting window. 


\begin{example}\label{ex:alwaysEventuallyZeroCount}
{\codesize
\begin{verbatim}
def alwaysEventuallyZeroCount = {
  type U = (RDD[Status], RDD[(String, Int)])
  val windowSize = 4   
  val (numBatches, maxHashtagLength) = (windowSize * 4, 8)
  
  // repeat hashtags a bit so counts are bigger than 1   
  val tweets = for {
    hashtags <- Gen.listOfN(6, hashtag(maxHashtagLength))
    tweets <- BatchGen.ofNtoM(5, 10, 
                addHashtag(Gen.oneOf(hashtags))(tweet(noHashtags=true)))
  } yield tweets
  val emptyTweetBatch = Batch.empty[Status]
  val gen = BatchGen.always(tweets, numBatches) ++ 
            BatchGen.always(emptyTweetBatch, windowSize*2)
  
  val alwaysEventuallyZeroCount = alwaysF[U] { case (statuses, _) =>
    val hashtags = getExpectedHashtagsForStatuses(statuses)
    laterR[U] { case (_, counts) => 
      val countsForStatuses = 
        hashtags
          .map((_, ()))
          .join(counts)
          .map{case (hashtag, (_, count)) => count}
      countsForStatuses should foreachRecord { _ == 0}
    } on windowSize*3
  } during numBatches
  
  forAllDStream(gen)(
    TweetOps.countHashtags(batchInterval, windowSize)(_))(
    alwaysEventuallyZeroCount)
}
\end{verbatim}
}
\end{example}

~\\
One example of the second kind of liveness properties, that use an implication in the body of an always, is the following property for \texttt{getTopHashtag}, that checks that if we superpose two generators, one for a random noise of hashtags that have a small number of occurrences (generator \texttt{tweets}), and another for a periodic peak of a random hashtag that suddenly has a big number of occurrences (generator \texttt{tweetsSpike}), then each time a peak happens then the corresponding hashtag eventually becomes the top hashtag. 

\begin{example}\label{ex:alwaysPeakImpliesEventuallyTop}
{\codesize
\begin{verbatim}
def alwaysPeakImpliesEventuallyTop = {
  type U = (RDD[Status], RDD[String])
  val windowSize = 2
  val sidesLen = windowSize * 2 
  val numBatches = sidesLen + 1 + sidesLen
  val maxHashtagLength = 8
  val peakSize = 20
  
  val emptyTweetBatch = Batch.empty[Status]
  val tweets =
    BatchGen.always(
      BatchGen.ofNtoM(5, 10, 
                     tweetWithHashtagsOfMaxLen(maxHashtagLength)), 
      numBatches)      
  val popularTweetBatch = for {
    hashtag <- hashtag(maxHashtagLength)
    batch <-  BatchGen.ofN(peakSize, tweetWithHashtags(List(hashtag)))
  } yield batch
  val tweetsSpike = BatchGen.always(emptyTweetBatch, sidesLen) ++
                         BatchGen.always(popularTweetBatch, 1) ++
                         BatchGen.always(emptyTweetBatch, sidesLen)
  // repeat 6 times the superposition of random tweets
  // with a sudden spike for a random hastag
  val gen = Gen.listOfN(6, tweets + tweetsSpike).map{_.reduce(_++_)}
 
  val alwaysAPeakImpliesEventuallyTop = alwaysF[U] { case (statuses, _) => 
    val hashtags = getExpectedHashtagsForStatuses(statuses)
    val peakHashtags = hashtags.map{(_,1)}.reduceByKey{_+_}
			          .filter{_._2 >= peakSize}.keys.cache()
    val isPeak = Solved[U] { ! peakHashtags.isEmpty }
    val eventuallyTop = laterR[U] { case (_, topHashtag) => 
      topHashtag must beEqualAsSetTo(peakHashtags)
    } on numBatches
    
    isPeak ==> eventuallyTop
  } during numBatches * 3 
                               
  forAllDStream(
    gen)(
    TweetOps.getTopHashtag(batchInterval, windowSize)(_))(
    alwaysAPeakImpliesEventuallyTop)
}
\end{verbatim}
}
\end{example}

The consume operator is also useful to define other types of properties like the following, that only uses consume and next as temporal operators, but that is able to express the basic condition for counting correctly and on time. 
It states that for any number of repetitions $n$ less or equal to the counting window size, and for any random word prefix, if we repeat the word prefix $n$ times then after the $n-1$ instants we will have a count of at least (to account for hashtags randomly generated twice) $n$ for all the hashtags in the first batch. Here we use \texttt{def next[T](times: Int)(phi: Formula[T])} that returns the result of applying next \texttt{times} times on the given formula.  

\begin{example}\label{ex:forallNumRepetitionsLaterCountNumRepetitions}
{\codesize
\begin{verbatim}
def forallNumRepetitionsLaterCountNumRepetitions = {
  type U = (RDD[Status], RDD[(String, Int)])
  val windowSize = 5
  val (numBatches, maxHashtagLength) = (windowSize * 6, 8)

  // numRepetitions should be <= windowSize, as in the worst case each
  // hashtag is generated once per batch before being repeated using
  // Prop.forAllNoShrink because sscheck currently does not support shrinking 
  Prop.forAllNoShrink(Gen.choose(1, windowSize)) { numRepetitions =>
    val tweets = BatchGen.ofNtoM(5, 10, 
                                 tweetWithHashtagsOfMaxLen(maxHashtagLength))      
    val gen = for {
      tweets <- BatchGen.always(tweets, numBatches)
      // using tweets as a constant generator, to repeat each generated
      // stream numRepetitions times
      delayedTweets <- PDStreamGen.always(tweets, numRepetitions)
    } yield delayedTweets

    val laterCountNumRepetitions = nextF[U] { case (statuses, _) =>
      val hashtagsInFirstBatch = getExpectedHashtagsForStatuses(statuses)
      // -2 because we have already consumed 1 batch in the outer nextF, and 
      // we will consume 1 batch in the internal now  
      next(max(numRepetitions-2, 0))(now { case (_, counts) =>
        val countsForHashtagsInFirstBatch = 
          hashtagsInFirstBatch
            .map((_, ()))
            .join(counts)
            .map{case (hashtag, (_, count)) => count}
        countsForHashtagsInFirstBatch should foreachRecord { _ >= numRepetitions }
      }) 
    }
    forAllDStream(
      gen)(
      TweetOps.countHashtags(batchInterval, windowSize)(_))(
      laterCountNumRepetitions)
  }
}
\end{verbatim}
}
\end{example}

\end{document}